\documentclass[preprint,3p,authoryear]{elsarticle}

\usepackage{amssymb}
\usepackage{amsmath}

\usepackage{xcolor}
\usepackage{algorithm}
\usepackage{algpseudocode}
\usepackage{hyperref}

\usepackage{subcaption}

\journal{Nuclear Physics B}

\begin{document}

\begin{frontmatter}

\title{Stochastic Modeling of Anisotropic Strength Surfaces from Atomistic Simulations} 

\author{A.\ Bonacci, J.E.\ Dolbow, J.\ Guilleminot}

\affiliation{organization={Mechanical Engineering and Materials Science, Duke University},city={Durham},
            postcode={27708}, 
            state={NC},
            country={USA}}

\begin{abstract}
This work develops a unified framework for inferring, representing, and statistically characterizing an anisotropic strength surface directly from molecular dynamics data. Large-scale tensile loading simulations are used to generate failure data across all principal stress ratios and loading orientations, facilitated by a data-driven mapping between imposed strain-rate tensors and resulting stresses. The orientation-dependent strength surface is then represented using a constrained parametric formulation in which the surface parameters vary smoothly with loading angle through a low-dimensional functional encoding. To deploy the framework, we specifically consider the case of monocrystalline graphene, which is a prototypical two-dimensional material that has been extensively characterized, both experimentally and computationally, in the literature. For defective graphene, multiple random realizations of vacancy defect distributions are used to construct a stochastic ensemble of angular strength surfaces. Because each anisotropic strength surface requires substantial atomistic sampling to construct, the resulting ensemble is inherently limited in size, motivating the use of compact encoding, dimensionality reduction, and probabilistic modeling to characterize strength variability. Dimensionality reduction via Principal Component Analysis reveals a condensed latent representation of the fitted, encoded surfaces, where a Gaussian mixture model is employed to capture defect-induced variability, including rare outlier behaviors arising from clustered vacancy defects. Sampling from this probabilistic model enables the generation of new, physically admissible strength surfaces and the construction of confidence intervals in both parameter space and stress space. Applied to graphene, the results show that the fracture strength exhibits pronounced anisotropy, weakens in the presence of vacancies, and presents strong variability governed primarily by defect density rather than defect type. 
\end{abstract}


\end{frontmatter}

\section{Introduction}

Predictive modeling of elastic brittle fracture requires an accurate description not only of how cracks propagate, but of the stress states at which they nucleate. To be truly predictive, modern fracture frameworks rely on three independent material properties: (1) elasticity, which governs reversible deformation; (2) fracture toughness, which dictates resistance to crack growth; and (3) strength, defined as the set of all critical stresses at which a material fractures when subjected to a monotonically increasing, spatially uniform, but otherwise arbitrary stress \citep{kumar_phase-field_2020, kumar_revisiting_2020}. Unlike elasticity or toughness, strength is not a fixed quantity such as a uniaxial tensile or compressive limit; instead, this collection of critical stress states forms a multidimensional boundary in stress space, known as the strength surface. Strength represents the macroscopic manifestation of microscopic defects, which gives rise to significant inherent stochasticity. Because of this intrinsically multidimensional and stochastic nature, strength remains the least understood and most challenging of the three properties to characterize experimentally or computationally. Despite these difficulties, a complete and accurate stochastic representation of the strength surface across loading conditions is essential for any framework seeking to predict when, and how, brittle materials fail.

Traditionally, efforts to characterize strength have relied on experimental approaches, where failure is observed under carefully controlled loading paths. While such experiments have yielded valuable insights \citep{broutman_effects_1970,ely_strength_1972,sato_fracture_1987}, they remain fundamentally limited. Multiaxial stress states are difficult to impose with precision, and due to the stochastic nature of fracture events, the number of tests required to obtain statistically meaningful strength surfaces is often prohibitively large. In many cases, direct experimental probing of the full failure envelope is simply not feasible.

The present work adopts an atomistic approach to address these challenges. By leveraging molecular dynamics (MD) simulations, wherein the atomic structure and interactions are prescribed, we can compute the full stress tensor at failure and aim to infer the complete strength surface of a material, including the stochastic variability introduced by defects. This enables the systematic probing of strength across all principal stress states, under controlled and repeatable conditions, in a way that complements and extends beyond the reach of experimental methods.

Graphene, a single atomic layer of sp\textsuperscript{2}-bonded carbon atoms arranged in a hexagonal lattice, provides a thoroughly-defined atomic structure that exhibits exceptional in-plane strength and stiffness---making it an ideal platform for developing and validating a model capable of inferring atomistic-scale strength. Despite these advantages, its mechanical response is highly sensitive to lattice orientation \citep{ni_anisotropic_2010, jiang_chirality-dependent_2017,  sutrakar_fracture_2021, qu_anisotropic_2022}. This motivates the definition of an orientation-dependent strength surface, allowing for the strength characterization to change based on lattice orientation. Atomic-scale vacancy defects also play a significant role in graphene's mechanical response, adding a stochastic element to its strength \citep{jing_effect_2012, hess_fracture_2016, gavallas_mechanical_2023}. These attributes call for the construction of a stochastic, anisotropic model of strength, one that captures the orientation-dependent behavior of the material as well as the inherent uncertainty arising from its defects. Understanding and quantifying this variability is essential for defining statistically robust design limits in anisotropic brittle materials, with graphene-based systems serving as a representative example.

Traditional MD studies have provided valuable insight into graphene’s fracture mechanisms under specific loading conditions \citep{akinwande_review_2017, kumar_mechanical_2021, torkaman-asadi_atomistic_2022}, yet these analyses were limited to a small number of orientations, defect configurations, and stress states. Moreover, these works emphasize general constitutive laws or qualitative failure behavior, rather than focusing on strength itself. As a result, they provide an incomplete view of the full anisotropic failure envelope. Additionally, the influence of random defect topologies introduces variability that deterministic approaches cannot describe. To move beyond these limitations, a framework is needed that can (i) generate strength data across all critical stresses at all loading orientations, (ii) account for the stochastic effects of defect morphology and density, and (iii) compress this information into a compact form suitable for coarser-scale modeling.

In this work, we develop such a framework to quantify and analyze stochastic, anisotropic strength, employing monocrystalline graphene as a model system. Using large-scale MD simulations, we utilize a data-driven mapping between imposed strain rates and resulting stresses to enable the generation of orientation-dependent failure data across the full range of principal stress states and loading directions. The failure envelope is then represented through an angularly dependent strength surface model in which the surface parameters are defined as smooth, periodic functions of orientation. Finally, we introduce a statistical inference procedure to characterize the variability in these angular strength surfaces caused by defect distributions. Because each surface realization is costly to obtain, the dataset is necessarily limited, motivating the use of Principal Component Analysis to extract a compact latent representation of the fitted surface parameters, followed by probabilistic modeling to characterize their variability. This enables the construction of a stochastic strength model capable of sampling new physically admissible surfaces and generating confidence intervals for the anisotropic strength response.

This unified approach builds a foundation aiming to bridge atomistic simulation, continuum modeling, and statistical characterization, constructing an uncertainty-aware strength surface primed for material design and analysis. While monocrystalline graphene serves as a well-defined model system in this study, the framework is agnostic to material choice and strength surface formulation, relying only on failure data across loading states. By linking defect statistics to anisotropic strength variability, this work establishes a physically grounded pathway toward multiscale models that incorporate defect-induced uncertainty---an essential step towards the reliable, predictive modeling of materials for engineering applications.

The remainder of this paper is organized as follows. Section \ref{sec:Methodology} presents the methodology, beginning with the molecular dynamics framework and system setup used to generate failure data. This is followed by the construction of strength surfaces---beginning with deterministic, orientation-fixed strength surfaces. Then, we work towards orientation-dependent representations, introducing angular encodings and physically admissible parameterizations. A stochastic element is introduced in the form of defects, where the fitting, dimensionality reduction, and probabilistic sampling strategies are discussed. Section \ref{sec:Results} presents the corresponding results, where we first examine the elastic and fracture response of pristine graphene. Next, we examine the stochastic effects induced by vacancy defects, delving into model decisions such as encoding strategies and the number of retained latent dimensions. We then analyze the resulting anisotropic strength surfaces and their statistical representations. The paper concludes with a discussion of the broader implications of the proposed framework and its potential extensions to other materials and length scales.

\section{Methodology}\label{sec:Methodology}
\subsection{Molecular Dynamics Framework}

Molecular dynamics simulations were performed on monocrystalline graphene sheets, using LAMMPS \citep{LAMMPS}. Each instance consisted of a $60 \times 60$ atom lattice, producing a rectangular domain with the armchair edge approximately 125~\r{A} in length and the zigzag edge approximately 72~\r{A}, which is comparable to the typical size of one graphene crystal \citep{wang_effects_2024}. This system size was selected as a compromise between computational cost and physical fidelity. In this work, we will analyze both pristine graphene sheets, and those with defects present. A large enough sheet is required to allow defects to be introduced in a statistically meaningful manner, with the selected domain size reflecting a practical balance between capturing relevant fracture behavior and limiting computational expense. Periodic boundary conditions were applied in-plane, with a sufficiently large vacuum spacing out-of-plane to approximate an infinite sheet and avoid spurious interactions across this boundary. In this context, both lattice dimensions must contain an even number of atoms; if one dimension contains an odd number, the hexagonal structure is not preserved across the periodic boundary, leading to artificial weakness at the edges. Simulations were advanced with a time step of 0.0005~ps, and the temperature was maintained at 273~K using a Nosé–Hoover thermostat \citep{hoover_canonical_1985} within the NVT ensemble. The carbon–carbon interactions were modeled using the AIREBO potential \citep{stuart_reactive_2000}.

As with any atomistic modeling approach, the fidelity of the inferred strength is fundamentally conditioned on the accuracy of the underlying interatomic potential chosen, particularly in the highly strained regime near fracture. This framework does not aim to introduce additional constitutive assumptions beyond those inherent to the molecular dynamics model; instead, it provides a systematic means of extracting and statistically characterizing strength behavior implied by the chosen potential. Consequently, the use of alternative potentials may be incorporated directly within the same workflow, without modification to the strength surface formulation or statistical inference procedure. 

\begin{figure}[ht!]
\centering
\includegraphics[width=0.6\linewidth]{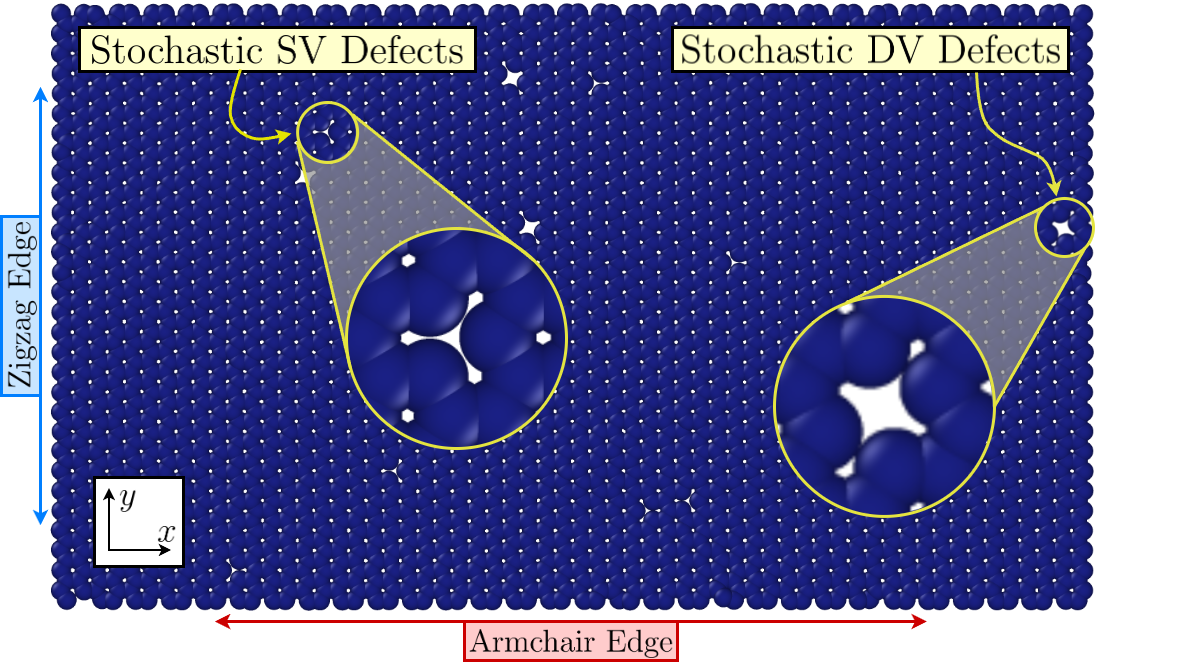}
\caption{\label{fig:system_vis} Schematic of a graphene sheet, with armchair and zigzag directions indicated, as well as examples of single vacancy (SV) and double vacancy (DV) defects. Visualized with the OVITO package \citep{ovito}.}
\end{figure}

Prior to stress testing, a pristine graphene system was relaxed from the initial rigid lattice configuration to a thermally equilibrated state. Random atomic velocities corresponding to 273~K were first assigned, after which the system was briefly stabilized under NVT conditions. The simulation was then allowed to evolve until the total energy reached equilibrium. The equilibrated configuration was saved and used as the baseline for all subsequent strength tests. In simulations involving defects, vacancies were introduced into the relaxed lattice at this stage. A validation study confirmed that different relaxation seeds did not affect the overall fracture behavior, aside from rare outlier cases at sheet edges. These atypical configurations were identified and excluded from the dataset to ensure consistency across simulations. A visualization of the system can be seen in Figure~\ref{fig:system_vis}, which also shows the stochastic vacancy defects considered in this work. These defects will be discussed in greater detail in Section~\ref{sec:stochastic_angular_ss}.

Mechanical loading was applied to impose a prescribed engineering strain rate tensor $\boldsymbol{\dot{\varepsilon}}$. This allowed for independent control of the normal strain rates $\dot{\varepsilon}_x$ and $\dot{\varepsilon}_y$, as well as the shear strain rate $\dot{\varepsilon}_{xy}$, while all other tensor components were fixed at zero. The maximum strain rate applied for any given component of the strain rate tensor was $\dot \varepsilon_{\text{max}} = 0.001$. This value was selected based on a convergence study identifying the largest strain rate that does not influence the observed fracture behavior, and is consistent with strain rates reported in prior molecular dynamics studies by \citet{jiang_mechanical_2014}. The global virial stress tensor $\boldsymbol{\sigma}$ was normalized by the initial volume of the graphene sheet, yielding a nominal stress measure that was extracted at specified time steps for subsequent analysis. Because of graphene’s anisotropy and the nonlinear nature of its response, the relationship between applied strain rate and resulting stress is not one-to-one; this point is discussed in Section~\ref{sec:Orient-dep-construct} and elaborated upon in \ref{app:nn_mapping}.

In MD simulations, fracture initiates through atomic bond breaking, but the failure of a small number of bonds does not necessarily produce a measurable change in the global stress response. To robustly identify system-level fracture events, we therefore define fracture as a sudden loss of stiffness subject to multiple consistency checks. Specifically, we require sufficiently long trajectories, a drop in the ergodic mean of at least one principal stress relative to a preceding baseline, the presence of a clear peak in that principal stress, and a post-peak low that exceeds a prescribed minimum drop threshold. Fracture is declared only if all checks are satisfied; otherwise, the simulation continues. This strategy is explained in more detail in~\ref{app:fracture_alg}.

\subsection{Orientation-Fixed Strength Surface Construction}\label{sec:Orientation-Fixed Strength Surface Construction}
\begin{figure}[ht!]
\centering
\includegraphics[width=0.7\linewidth]{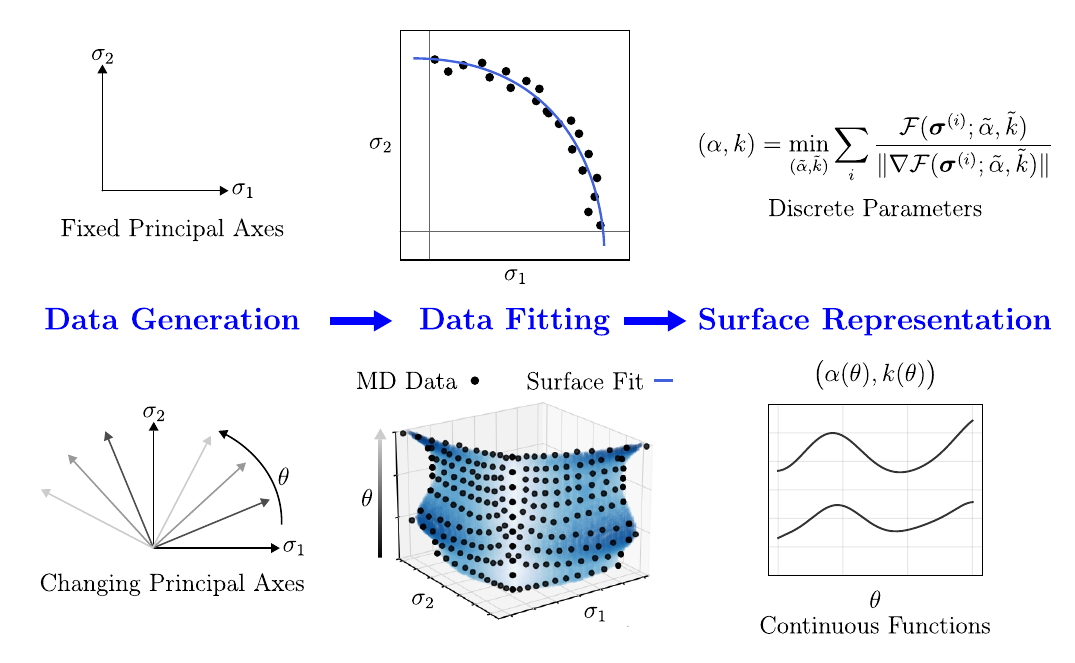}
\caption{\label{fig:workflow_1} Schematic overview of the strength surface construction workflow. The top row represents a conventional plane stress strength surface obtained from MD simulations in which the principal stress axes are fixed relative to the lattice. In this setting, failure data lie on a two-dimensional envelope in $(\sigma_1, \sigma_2)$ space and are represented by a strength surface parameterized by constant coefficients $(\alpha, k)$. The bottom row represents generalized data generation in which the principal axes rotate relative to the lattice, introducing an explicit angular dependence $\theta$. The resulting MD failure data populate a three-dimensional space $(\sigma_1, \sigma_2, \theta)$, and fitting yields orientation-dependent strength parameters $\alpha(\theta)$ and $k(\theta)$, represented as continuous functions of loading angle. The fixed-axis strength surface is recovered as a slice of this angularly dependent surface at constant $\theta$, illustrating the natural increase in model complexity required to capture anisotropic fracture behavior.}
\end{figure}

To construct a strength surface, we require MD failure data spanning a range of principal stress ratios $\sigma_2/\sigma_1 \in [0,1]$, where $\sigma_1 \ge \sigma_2 \ge \sigma_3$ denote the principal stresses---defined as the eigenvalues of the stress tensor ordered by magnitude---with $\sigma_1$ corresponding to the maximum principal stress at failure. Under the plane stress conditions considered here, $\sigma_1$ and $\sigma_2$ fully characterize the relevant stress states. This range of principal stress ratios ensures full coverage across the entire tensile strength domain, from pure uniaxial tension ($\sigma_2/\sigma_1 = 0$) to pure biaxial tension ($\sigma_2/\sigma_1 = 1$). The two-dimensional nature of graphene insists tensile-only loading conditions. If in-plane compression is applied to the graphene sheet, it will buckle out of plane, causing nonuniform stress fields that violate the assumption of spatially uniform loading required to define a strength surface. For the simplest loading scheme, we set $\dot \varepsilon_{xy} = 0$, which fixes the principal axes to coincide with the $x$–$y$ coordinate frame signified in Figure~\ref{fig:system_vis}. Here, we can apply loads where the dominant principal direction is aligned with the $x$-axis (0\textdegree{}) by setting $\dot{\varepsilon}_{x} = \dot{\varepsilon}_{\max}$ while varying $\dot{\varepsilon}_y$ between $[0, \dot{\varepsilon}_{\max}]$. This produces a set of failure points that resemble a strength envelope in principal stress space, which can be visualized by the points in the data fitting section of the top row of Figure~\ref{fig:workflow_1}. 
We can repeat this process, but instead set $\dot{\varepsilon}_y = \dot{\varepsilon}_{\max}$, and vary $\dot{\varepsilon}_x$ between $[0, \dot{\varepsilon}_{\max}]$. This creates MD failure points in which the dominant principal direction aligns with the $y$-axis (90\textdegree). 

Because of the hexagonal lattice, applying a strain rate consisting of simple uniaxial tension along the $y$-axis does not yield a purely uniaxial stress state in the $y$-direction; some stress is redistributed to the $x$-direction. Accordingly, we must apply slight compressive strain in the $x$-direction to achieve the uniaxial result. This necessitates varying $\dot \varepsilon_x$ between $[-0.1\dot \varepsilon_{\max}, \dot \varepsilon_{\max}]$ to ensure all stress ratios including pure uniaxial tension are represented, without inducing out-of-plane buckling or bending.

This process yields the MD failure points along a strength surface for both the armchair (0\textdegree{}) and zigzag (90\textdegree{}) directions. For the 0\textdegree{} orientation, eleven failure points were generated to span principal stress ratios from uniaxial to biaxial tension. The 90\textdegree{} zigzag case had twelve failure points, where an additional point at $\dot{\varepsilon}_x = -0.1\dot{\varepsilon}_{\max}$ was included to capture the true uniaxial response.

To reduce the dimensionality of this data and provide a continuous description, we fit the failure points with a strength surface model defined by \citet{drucker_soil_1952}:
\begin{equation}{\label{eqn:DP}}
\mathcal{F}(\boldsymbol{\sigma}) = \sqrt{J_2(\boldsymbol{\sigma})} + \alpha I_1(\boldsymbol{\sigma}) - k = 0.
\end{equation}
Here, $I_1(\boldsymbol{\sigma})$ and $J_2(\boldsymbol{\sigma})$ are the first and second invariants of the stress tensor $\boldsymbol{\sigma}$, and $\alpha$ and $k$ are material parameters that control the shape and location of the surface. The curve in the top row of Figure~\ref{fig:workflow_1} represents a visualization of this surface in stress space, where the intersection with the axes represents the uniaxial tensile strength, and the point where $\sigma_1 = \sigma_2$ represents the biaxial tensile strength. The resulting $(\alpha, k)$ parameters provide a convenient way to characterize a strength surface.\footnote{The two-parameter Drucker–Prager model adopted here serves primarily as a convenient low-dimensional representation. The broader framework is agnostic to the particular choice of strength surface and may accommodate alternative formulations with minimal modification.}

To ensure a physically admissible strength surface, explicit bounds on $\alpha$ and $k$ must be enforced. Rearranging Eq.~\eqref{eqn:DP} in terms of the stress invariants for the uniaxial tensile strength $\sigma_{ts}$ and the biaxial tensile strength $\sigma_{bs}$ yields:
\begin{equation}
    \sigma_{ts} = \frac{3k}{\sqrt{3} + 3\alpha}, \qquad \sigma_{bs} = \frac{3k}{\sqrt{3} + 6\alpha}.
\end{equation}
Both strengths must remain positive; otherwise the surface would imply a nonphysical infinite strength, contrary to MD and physical observations. This requirement leads to the conditional bounds:
$$
    \text{if } k > 0 \text{ then } \alpha > -\sqrt{3}/6, \qquad
\text{if } k < 0, \text{ then } \alpha < -\sqrt{3}/3.
$$
For this particular application on graphene, it was observed that all fitted $k$ values are positive. While negative $k$ values are theoretically possible, they were not observed here. Consequently, we restrict the admissible region to 
\begin{equation}
    k > 0 \quad \text{and} \quad \alpha > -\sqrt{3}/6,
\end{equation}
ensuring both uniaxial and biaxial tensile strengths remain finite. It is important to note that the combination of $\alpha < 0$ and $k > 0$ would imply inadmissible hydrostatic strengths in a full three-dimensional geometry. However, because this study is conducted under plane stress conditions, admissibility is instead governed by the biaxial tensile strength rather than the hydrostatic strength. Accordingly, the lower bound on $\alpha$ is shifted from $0$---which arises from the hydrostatic constraint---to $-\sqrt{3}/6$, which ensures positive biaxial tensile strength and therefore physical admissibility in the present framework.

This approach collapses the MD failure data into two parameters $(\alpha, k)$ that will serve as the foundation (i.e., input features) for subsequent statistical analysis in the defective graphene cases.

\subsection{Anisotropic Strength Surface Construction}\label{sec:Orient-dep-construct}
To probe orientations not aligned with the armchair or zigzag axes, we introduce shear strain rates that rotate the principal directions relative to the lattice. This corresponds with the bottom row of Figure~\ref{fig:workflow_1}. In this regime, the relationship between the components of the imposed in-plane strain-rate tensor, collected in $\boldsymbol{\dot\varepsilon} = \{\dot \varepsilon_x, \dot \varepsilon_y, \dot \varepsilon_{xy} \}$, and that of the resulting stress tensor gathered in $\boldsymbol \sigma = \{\sigma_x, \sigma_y, \sigma_{xy}\}$, becomes nonlinear and non-unique. To handle this, we constructed a data-driven mapping between strain-rate and stress tensors using molecular dynamics data, enabling us to predict the stress response for arbitrary loading combinations. The forward mapping
\begin{equation}
    \boldsymbol \sigma = f(\boldsymbol{\dot \varepsilon}),
\end{equation}
was learned using a neural-network regression model, and an inverse optimization routine was then employed to identify the strain-rate tensor corresponding to any desired principal stress ratio and orientation. A detailed description of the neural network architecture, training procedure, and inverse mapping workflow is provided in~\ref{app:nn_mapping}, where the implementation and validation results are reported.

\subsubsection{Principal Direction Calculation}\label{sec:Principal Direction Calculation}
To generate MD failure data at intermediate loading angles, strain rate tensors corresponding to arbitrary target orientations were obtained using the neural-network mapping described in~\ref{app:nn_mapping}, enabling us to probe principal stress ratios spanning $[0, 1]$ and angles of the dominant principal direction spanning $[0, 90]$\textdegree{} with respect to the bottom armchair edge of the lattice. The orientation of the dominant principal direction was determined from the eigen-decomposition of the stress tensor. For each time step, the stress tensor $\boldsymbol\sigma$ was entered into the eigenvalue problem
\begin{equation}
    \boldsymbol \sigma \mathbf v^{(i)} = \sigma_i \mathbf v^{(i)},
\end{equation}
which was solved to obtain the principal stresses $\sigma_1$, $\sigma_2$, and $\sigma_3$, and the corresponding principal directions $\mathbf{v}^{(1)}$, $\mathbf{v}^{(2)}$, and $\mathbf{v}^{(3)}$. The eigenvalues were sorted from largest to smallest, ensuring that $\sigma_1$ and $\mathbf{v}^{(1)}$ correspond to the dominant principal stress and its orientation.

The loading angle was calculated by projecting the dominant eigenvector $\mathbf{v}^{(1)}$ into the $x$-$y$ plane and comparing with the $x$-axis:
\begin{equation}
    \theta_g = \arctan2 (v_y^{(1)}, v_x^{(1)}),
\end{equation}
where $v_x^{(1)}, v_y^{(1)}$ are the in-plane components of $\mathbf v^{(1)}$. Here, $\theta_g$ represents the angle of the dominant eigenvector with respect to the global coordinate frame. The angle was then converted to degrees and folded into the interval $[0, 90]$\textdegree{} using the symmetry relation $\theta \mapsto \min(\theta, \; 180^\circ - \theta)$.

While the procedure above provides the angle of the dominant principal stress with respect to the simulation box, what we are truly interested in is the angle of loading relative to the lattice. When shear components are applied, the lattice itself rotates during deformation, and failure behavior depends on this lattice-referenced orientation. If this effect is ignored, the calculated loading angle would reflect orientation only with respect to the simulation frame and not the evolving material frame. To correct for this, we employed the polar decomposition of the deformation gradient. The deformation gradient $\mathbf{F}$ was constructed from the prescribed strain rates at simulation time $t$, 
\begin{equation}
    \mathbf F = \begin{bmatrix} 1 + \dot \varepsilon_x(t) && \dot \varepsilon_{xy}(t) && \dot \varepsilon_{xz}(t) \\ 0 && 1 + \dot \varepsilon_y(t) && \dot \varepsilon_{yz}(t) \\ 0 && 0 && 1 + \dot \varepsilon_z(t) \end{bmatrix},
\end{equation}
and then decomposed into a pure rotation and a pure stretch,
\begin{equation}
    \mathbf F = \mathbf{RU},
\end{equation}
where $\mathbf{R}$ is a proper orthogonal tensor ($\det \mathbf{R} = 1$, $\mathbf{R}^T \mathbf{R} = \mathbf{I}$) representing the rigid-body lattice rotation, and $\mathbf{U}$ is a symmetric positive-definite tensor representing the stretch. The rotation tensor $\mathbf{R}$ isolates the change of lattice orientation due to shear, and thus provides the reference frame required for an accurate measurement of the angle $\theta$.

In practice, $\mathbf{R}$ was extracted using the right-polar decomposition, and the instantaneous lattice rotation was computed directly from the first column of $\mathbf{R}$. The rotation angle was then expressed in degrees as
\begin{equation}
    \theta_r=\arctan2 (R_{21}, R_{11}).
\end{equation}
This gives the orientation of the lattice $x$ axis, which was originally the 0\textdegree{} armchair edge, compared to the global $x$-axis. By subtracting this lattice rotation from the  global dominant principal direction angle computed earlier, we obtain the effective loading angle relative to the lattice;
\begin{equation}
    \theta = \theta_g - \theta_r.
\end{equation}
This adjustment ensures that the orientation-dependent strength surface accurately captures the material anisotropy by measuring loading directions in the material frame, independent of any rigid-body rotation induced by shear.

\subsubsection{Anisotropic Strength Surface Modeling}\label{sec:Anisotropic Strength Surface Modeling}

Utilizing the corrected orientation angle, each MD failure event can now be represented in a three-dimensional space spanned by the two principal stresses $(\sigma_1,\sigma_2)$ and the effective loading angle $\theta$, as visualized by the three-dimensional plot at the bottom of Figure~\ref{fig:workflow_1}. This representation exposes the full anisotropy of the material: instead of lying on a single two-dimensional strength envelope, the failure points trace out a continuous angular dependence---essentially forming a wavy three-dimensional surface---as the loading direction rotates from 0\textdegree{} to 90\textdegree. This calls for the expansion of the strength surface definition provided in Eq.~\eqref{eqn:DP}, now allowing $\alpha$ and $k$ to vary explicitly with $\theta$:
\begin{equation}\label{eqn:angular_dp}
    \mathcal{F}(\boldsymbol{\sigma}, \theta) = \sqrt{J_2(\boldsymbol{\sigma})} + \alpha(\theta) I_1(\boldsymbol{\sigma}) - k(\theta) = 0.
\end{equation}

A useful way to interpret this structure is to examine cross-sections at fixed $\theta$. Each such slice resembles a conventional 2D plane stress strength surface, while the collection of slices across all angles reveals how the envelope evolves smoothly with orientation. This provides a direct link between lattice orientation and fracture strength, setting the foundation for a parametric model of angular dependence.

We now must define $\alpha(\theta)$ and $k(\theta)$ in a way that is both mathematically tractable and physically consistent. To this end, we introduce an angular reparameterization $\omega(\theta)$ that enables a smooth, periodic functional encoding of the orientation dependence, with the specific form of $\omega$ chosen to reflect the underlying material symmetry.
\begin{itemize}
    \item For pristine graphene, the lattice possesses sixfold rotational symmetry, resulting in a mechanical response that is periodic every 60\textdegree. This periodicity reflects the equivalence of armchair orientations at $\theta = 0$\textdegree{} and 60\textdegree, and zigzag orientations at $\theta = 30$\textdegree{} and 90\textdegree. To preserve this symmetry, we define $\alpha$ and $k$ as functions of the angular coordinate
    \begin{equation}
    \omega = \frac{2\pi\theta}{60}, 
    \end{equation}
    ensuring that the fitted functions repeat exactly over the fundamental period of the hexagonal lattice.
    \item When defects are introduced, the rotational periodicity observed in pristine graphene no longer holds. The random distribution and orientation of vacancy defects break the inherent sixfold symmetry, meaning that the angular dependence of strength can no longer be assumed to repeat every 60\textdegree. To account for this loss of symmetry, we instead introduce the reparameterized angular coordinate
\begin{equation}
    \omega = \frac{2\pi\theta}{180},
\end{equation}
which maps the physical loading angle $\theta \in [0, 180]$\textdegree{} onto a $2\pi$-periodic domain. Although the introduction of defects breaks the sixfold lattice symmetry, the stress response itself remains naturally periodic over 180\textdegree, since tension applied in a given direction is mechanically equivalent to tension applied 180\textdegree{} opposite that direction. The use of $\omega$ therefore preserves the physical periodicity of the loading response while providing a convenient parameterization for basis expansions. 
\end{itemize}
The above choices are not specific to graphene: different crystalline symmetries, material classes, or loading conventions may naturally motivate alternative periodic mappings, without altering the structure of the strength surface representation introduced here. In principle, one could parameterize the angular dependence directly in terms of $\theta$ (i.e., by taking $\omega=\theta$); more generally, $\omega$ may be viewed as any smooth, one-to-one reparameterization of $\theta$ selected to reflect the relevant symmetry or periodicity of the material system under consideration. 

To construct a flexible yet physically admissible representation of the angular strength surface, we adopt a two-step formulation. We first introduce unconstrained auxiliary functions that admit a convenient finite-dimensional encoding, and then map these functions to physically meaningful strength parameters through a smooth transformation that enforces admissibility.

Specifically, we introduce auxiliary transformation functions $z_\alpha(\omega)$ and $z_k(\omega)$, defined on the periodic angular coordinate $\omega$. These functions are unconstrained and serve as an intermediate representation in which the angular dependence can be efficiently encoded. Physical strength-surface parameters are then obtained by applying a smooth, differentiable transformation that enforces the lower-bound constraints identified in Section~\ref{sec:Orientation-Fixed Strength Surface Construction}. In particular, we employ the softplus function,
\begin{equation}\label{eqn:physical_transformation}
\tilde \alpha(\omega) = -\tfrac{\sqrt{3}}{6} + \ln\big(1 + e^{z_\alpha(\omega)}\big), \qquad
\tilde k(\omega) = \ln\big(1 + e^{z_k(\omega)}\big),
\end{equation}
where the physical strength-surface parameters are obtained by composition:
\begin{equation}\label{eqn:reparameterize}
    \alpha(\theta) = \tilde \alpha(\omega(\theta)), \qquad k(\theta) = \tilde k(\omega(\theta)).
\end{equation}
This transformation guarantees that $\alpha > -\sqrt{3}/6$ and $k > 0$ for all $\theta$, ensuring that all reconstructed strength surfaces remain physically admissible across the full angular domain. In effect, $\tilde\alpha$ and $\tilde k$ serve as parametric representations in the periodic coordinate $\omega$, while $\alpha(\theta)$ and $k(\theta)$ remain functions of the physical loading angle.

Having established the physical mapping, we now specify how the auxiliary functions $z_\alpha(\omega)$ and $z_k(\omega)$ are represented. To capture the angular dependence in a compact form while preserving periodicity, each auxiliary function is expressed as a finite expansion in a chosen set of basis functions $\{\phi_j(\omega)\}_{j=0}^{n-1}$ with corresponding coefficients $z_j$:
\begin{equation}\label{eqn:fourier}
z_\alpha(\omega) = \sum_{j=0}^{n-1} z_j^{(\alpha)}\,\phi_j(\omega), \qquad
z_k(\omega) = \sum_{j=0}^{n-1} z_j^{(k)}\,\phi_j(\omega).
\end{equation}
Here, the basis functions $\phi_j(\omega)$ define the chosen encoder, which maps the continuous angular dependence into an $n$-dimensional coefficient space for $\alpha$ and $k$ respectively, resulting in a coefficient space that is $2n$-dimensional. 

The coefficients $\mathbf{z}_{\alpha}$ and $\mathbf{z}_{k}$ are determined by minimizing the least-squares misfit between $N$ molecular dynamics failure data points and the angular strength surface model:
\begin{equation}\label{eqn:minimization}
\mathbf z = \text{argmin}_{\tilde{\mathbf z}}~\mathcal{L}(\tilde{\mathbf z}),
\end{equation}
where $\mathbf z := (\mathbf z_\alpha^\top,\ \mathbf z_k^\top)^\top$ represents the vertical concatenation of column vectors, and $\tilde{\mathbf z}$ is a dummy minimization variable. The regularized loss function is defined as
\begin{equation}\label{eq:Ldef}
\mathcal{L}(\tilde{\mathbf z}) = \sum_{i=1}^N \big(r^{(i)}(\tilde{\mathbf z})\big)^2 + \lambda \frac{||\tilde{\mathbf z}||^2}{2},
\end{equation}
with
\begin{equation}\label{eq:res_def}
r^{(i)}(\tilde{\mathbf z}) = \frac{\mathcal{F}^{(i)}(\tilde{\mathbf z})}{||\nabla \mathcal{F}^{(i)}(\tilde{\mathbf z})||}
\end{equation}
and
\begin{equation}
\mathcal{F}^{(i)}(\tilde{\mathbf z})
:= \sqrt{J_2(\boldsymbol{\sigma}^{(i)})}
+ \alpha(\theta^{(i)};\tilde{\mathbf z}_\alpha)\, I_1(\boldsymbol{\sigma}^{(i)})
- k(\theta^{(i)};\tilde{\mathbf z}_k).
\end{equation}
Here $\boldsymbol{\sigma}^{(i)}$ and $\theta^{(i)}$ represent the \textit{i}th MD failure point. For notational clarity, the dependence of $\mathcal F^{(i)}(\tilde{\mathbf z})$ on these quantities is not written explicitly. The quadratic regularization term in Eq.~\eqref{eq:Ldef}, corresponding to standard Tikhonov regularization, controls the magnitude of the coefficients, promoting smooth angular functions and ensuring that the resulting coefficient space remains well-conditioned enough for meaningful statistical analysis in later sections. The regularization parameter $\lambda$ is tuned to balance these two objectives---restricting overgrowth of the coefficients without degrading the quality of the fit. This is discussed in greater detail in Section~\ref{sec:angular_ss_defective} surrounding Figure~\ref{fig:convergences}. Normalization by the gradient in the residual (see Eq.~\eqref{eq:res_def}) ensures that it represents the orthogonal geometric distance from the stress point to the failure surface, rather than the raw value of the implicit function $\mathcal F$. This prevents artificial weighting of regions where $\mathcal F$ is naturally steep or shallow, and yields a fit that is physically meaningful at the interface $\mathcal F = 0$.

The solution to Eq.~\eqref{eqn:minimization} provides a finite-dimensional encoding of the full angular strength surface. The formulation is agnostic to the choice of basis, making it readily extendable to various functional representations/encoders. Although polynomial splines and other bases can be considered, this work utilizes a truncated Fourier basis for $\phi_j(\omega)$, taking advantage of its natural periodicity and balance between smoothness, expressiveness, and generalizability.

The choice of Fourier basis functions motivates highlighting the relationship between the Fourier order $n_{\text{FO}}$ and the number of basis functions $n$:
\begin{equation}
    n = 2n_{\text{FO}}+1.
\end{equation}
Note that because both $\alpha(\theta)$ and $k(\theta)$ are represented with their own Fourier expansions, the total number of basis functions used in the angular strength surface is $2n$.

To determine an appropriate Fourier order, we increase $n_{\text{FO}}$ until the distribution of the normalized root-mean-square error (NRMSE), computed once per $\mathbf z$, stabilizes. The metric is defined as
\begin{equation}
    \text{NRMSE} = \frac{1}{\bar{\sigma}_1}\sqrt{\frac{1}{N} \sum_{i=1}^N \bigg(\frac{\mathcal{F}^{(i)}(\mathbf z)}{||\nabla \mathcal{F}^{(i)}(\mathbf z)||}\bigg)^2},
\end{equation}
where $\bar \sigma_1$ denotes the mean maximum principal stress across all MD simulations in the dataset of interest. In practice, full convergence of the NRMSE is neither expected nor necessary. Because the procedure is entirely data-driven, increasing $n_{\text{FO}}$ eventually introduces spurious high-frequency modes in the Fourier expansion, particularly in regions with low data density. These areas contribute negligibly to the loss, allowing higher-order series to contort in unrealistic ways to match isolated clusters of points. This behavior represents overfitting rather than improved physical fidelity. Consequently, a combination of NRMSE trends and qualitative agreement was used to identify an $n_{\text{FO}}$ that minimizes the normalized error while maintaining a physically reasonable angular dependence.

\subsubsection{Stochastic Anisotropic Strength Surface Characterization}\label{sec:stochastic_angular_ss}
Introducing defects into the lattice adds a stochastic component to the orientation-dependent strength surface. Here, we will focus on varying defect configurations, all with the same underlying defect density. Defect density is quantified by the fraction of atoms removed relative to the total atom count in the domain. Specifically, we examine three defect configurations with their labels (in parentheses):

\begin{enumerate}
    \item 0.5\% Single Vacancy defects (SV);
    \item 0.5\% Double Vacancy defects (DV);
    \item Mixed case with 0.25\% SV + 0.25\% DV defects (MX).
\end{enumerate}

A single vacancy defect is defined as the removal of a single atom in the graphene sheet, and a double vacancy defect is defined as the removal of two adjacent atoms \citep{jing_effect_2012}. Because each DV defect removes twice as many atoms as an SV defect, the 0.5\% DV case contains half as many individual defects as the 0.5\% SV case, while maintaining the same total number of atoms removed.

For each defect configuration, multiple random seeds generate distinct strength realizations of the defect distribution. Each realization yields its own set of MD failure points in $(\sigma_1, \sigma_2, \theta)$ space, from which an angular strength surface can be fit. As described in Section~\ref{sec:Anisotropic Strength Surface Modeling}, this yields a corresponding set of coefficients $\mathbf{z}$, which fully describe the angular variation of $\alpha(\theta)$ and $k(\theta)$. Collectively, these coefficient vectors form a dataset capturing the stochastic variability of the angular strength surface across many defect realizations and configurations.

While each $\mathbf{z}$ is $2n$-dimensional, some of its components are correlated or contribute negligibly to the overall variation. To obtain a more compact representation, we apply Principal Component Analysis (PCA) \citep{jolliffe_principal_2002} to the set of $\mathbf{z}$ vectors, which identifies the dominant eigenspace of the covariance matrix and provides an orthogonal basis in which the data can be efficiently projected into a reduced latent space. To determine how many components to retain, we consider the standard cumulative explained variance,
\begin{equation}
C(d) = \frac{\sum_{i=1}^{d} \Lambda_i}{\sum_{j = 1}^{2n} \Lambda_j},
\end{equation}
where $\Lambda_i$ denotes the \textit{i}th eigenvalue of the covariance matrix, and $d$ represents the number of principal components retained (assuming ordered eigenvalues). The corresponding recovery error,
\begin{equation}
E(d) = 1 - C(d),
\end{equation}
measures the fraction of total variance not captured when retaining $d$ components. The reduced dimensionality is selected by specifying a recovery-error tolerance $E_{\text{thresh}}$ and choosing the smallest integer $d$ such that $E(d) \leq E_{\text{thresh}}$. In practice, $E_{\text{thresh}}$ is chosen to ensure that the main features (e.g., data concentration) of the strength surface are properly captured. In this reduced latent PCA space, each realization is represented by a coordinate vector
\begin{equation}
    \boldsymbol{\eta} = [\eta_1, \eta_2, \ldots, \eta_d].
\end{equation} 
Inspection of the projected data shows that the latent variables exhibit a multimodal structure, driven by certain irregular strength responses. To faithfully represent this structure, we employ a Gaussian mixture model (GMM) \citep{bishop_pattern_2006} in the latent space. Specifically, the latent variable distribution is represented through a finite mixture of $M$ Gaussian components,
\begin{equation}\label{eqn:gmm}
    \boldsymbol{\eta} \sim \sum_{m=1}^M \pi_m \, \mathcal{N}(\boldsymbol{\mu}_m, \mathbf{\Sigma}_m),
\end{equation}
where $\{\pi_m\}_{m = 1}^M$ are the component weights satisfying $\sum_m \pi_m = 1$, and $\boldsymbol{\mu}_m$ and $\boldsymbol{\Sigma}_m$ denote the mean vector and covariance matrix of the $m$-th Gaussian component. The Gaussian mixture representation provides a flexible statistical model for the latent-space distribution. Selecting the number of mixture components, however, presents a well-known challenge. Standard model-selection criteria such as AIC, BIC, AICc, and likelihood-ratio tests \citep{burnham2002} were evaluated, but none produced meaningful guidance due to the intrinsic mismatch between the high dimensionality of the latent space and the limited number of realizations available. 

Alternatively, model selection can be guided by physical interpretation and parsimony (e.g., seeking the simplest mixture structure capable of adequately representing the variability observed in the MD-derived strength surfaces). For many defect configurations, the latent data cluster tightly enough that a single Gaussian component is sufficient (see Section~\ref{sec:angular_ss_defective}). For some scenarios involving pronounced defect-induced anisotropy or outlier strength surfaces, the use of a GMM with $M > 1$ becomes necessary. In our numerical experiments, the number of components $M$ was selected through qualitative agreement in terms of latent-space structure and evaluation of the model’s ability to reconstruct strength surfaces consistent with those produced directly from MD.

This modeling approach reflects a fundamental challenge: generating even a single angular strength surface requires substantial computational effort, making large datasets unattainable. As a result, statistical modeling must rely on prior representations that regularize over a limited number of samples, whether obtained from MD simulations or future experimental workflows. The proposed framework combining functional encoding, statistical reduction, and the Gaussian mixture model therefore strikes a balance between expressive power and feasibility, providing a statistically-sound and physically-interpretable means of characterizing the stochastic variability of orientation-dependent strength. Once this probabilistic model is defined, we can efficiently generate new physically-admissible strength surfaces through the following workflow (where the hat notation represents a sampled value):
\begin{enumerate}
    \item \textbf{Sampling:} Draw new latent vectors $\hat{\boldsymbol{\eta}}$ from the calibrated GMM (defined in Eq.~\eqref{eqn:gmm}).
    \item \textbf{Inverse PCA:} Map each sampled vector back to the full $2n$-dimensional coefficient space via the inverse PCA transformation, yielding new $\hat{\mathbf{z}}$ vectors.
    \item \textbf{Auxiliary reconstruction:} Substitute the sampled $\hat{\mathbf{z}}$ into the auxiliary transformation definitions of $\hat z_\alpha(\omega)$ and $\hat z_k(\omega)$ from Eq.~\eqref{eqn:fourier} (in this case, the Fourier transformation).
    \item \textbf{Physical transformation:} Apply the transformations in Eq.~\eqref{eqn:physical_transformation}, then apply the reparameterization in Eq.~\eqref{eqn:reparameterize} to recover bounded, physically admissible functions $\hat \alpha(\theta)$ and $\hat k(\theta)$ for all angles of loading.
\end{enumerate}

\begin{figure}[ht!]
\centering
\includegraphics[width=0.6\linewidth]{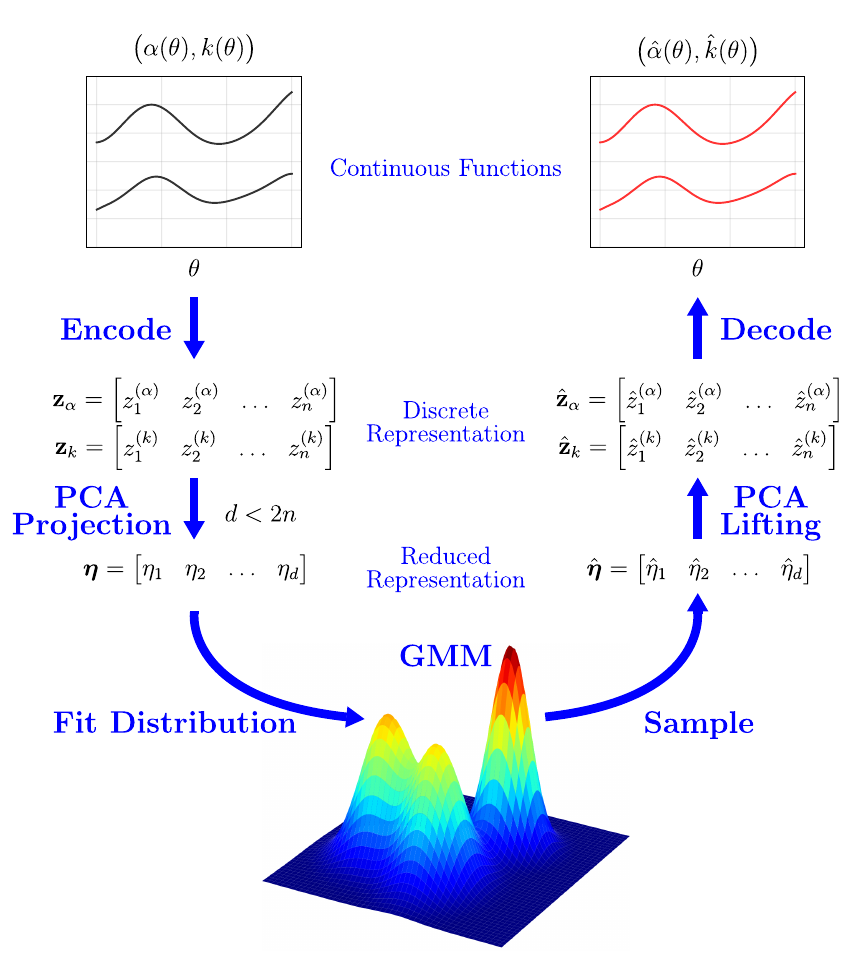}
\caption{\label{fig:workflow_2} Schematic of the stochastic modeling and sampling framework for physically admissible orientation-dependent strength surfaces. Starting from fitted continuous strength parameters $\alpha(\theta)$ and $k(\theta)$, an encoder maps the functions to a discrete coefficient vector $\mathbf z$ (here realized via a truncated Fourier representation). Principal Component Analysis (PCA) is then applied to $\mathbf z$ to obtain a reduced-order latent representation $\boldsymbol \eta$ with dimension $d < 2n$. The distribution of latent variables is modeled using a Gaussian mixture model (GMM), enabling probabilistic sampling in the reduced space. Sampled latent vectors $\hat{\boldsymbol \eta}$ are  lifted back to the full coefficient space via inverse PCA and decoded to recover continuous parameter functions $\big(\hat{\alpha}(\theta), \hat{k}(\theta)\big)$. The encoding, dimensionality reduction, and decoding steps collectively ensure that all sampled strength surfaces remain smooth and physically admissible.}
\end{figure}
Figure~\ref{fig:workflow_2} visualizes the proposed workflow, starting with the parametric $\big(\alpha(\theta), k(\theta)\big)$ functions from Figure~\ref{fig:workflow_1}, visualizing the encoding, reduced-order representation, and latent space fitting, as well as the sampling technique. By repeatedly sampling from the latent Gaussian mixture model, we generate an ensemble of angular strength surfaces consistent with the observed MD-derived variability. From this ensemble, various probabilistic quantities, such as confidence intervals and mean functions, can be computed for both $\alpha(\theta)$ and $k(\theta)$ across the full orientation domain. These statistics provide a quantitative picture of how defects introduce anisotropic variability strength, offering a direct pathway toward uncertainty-aware models of fracture behavior informed by nanoscale simulations.

In addition to the parameter-space representation, strength statistics can be computed directly in physical stress space. For any prescribed loading angle $\theta_0$, samples of $\alpha(\theta_0)$ and $k(\theta_0)$ can be mapped through the strength criterion defined in Eq.~\eqref{eqn:DP} to yield distributions of surfaces. This enables the construction of confidence intervals and probabilistic bounds directly in $(\sigma_1,\sigma_2)$, furnishing uncertainty estimates for specific loading directions. A fully continuous representation in $(\sigma_1,\sigma_2,\theta)$ space is also attainable; however, such three-dimensional visualizations tend to obscure rather than clarify the anisotropic trends, and thus the parameter-based and angle-sliced stress-space intervals provide the most interpretable and practically relevant summaries. By examining the angular dependence of the variance in $\alpha(\theta)$ and $k(\theta)$, one can identify orientations that are statistically more reliable or more defect-sensitive. Conditioning the latent-space distributions on defect class or density further allows the contributions of different defect types to be disentangled, clarifying whether defect morphology meaningfully alters the anisotropic response or whether density alone governs the behavior. In this way, the statistical tools developed here move beyond descriptive surface fitting, providing a quantitative basis for assessing the relative importance of orientation, defect structure, and defect density in governing stochastic fracture strength.

\section{Results}\label{sec:Results}
\subsection{\label{sec:Pristine Graphene}Analysis of Pristine Graphene}
\begin{figure}[!htb]
    \centering
    \begin{subfigure}[c]{0.32\linewidth}
        \centering
        \includegraphics[width=\linewidth]{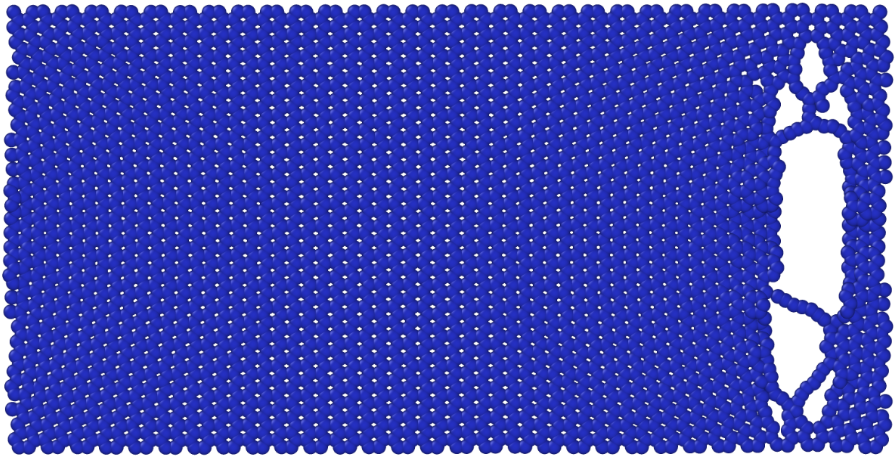}
        \caption{}
        \label{fig:uni_x}
    \end{subfigure}
    \hfill
    \begin{subfigure}[c]{0.32\linewidth}
        \centering
        \includegraphics[width=\linewidth]{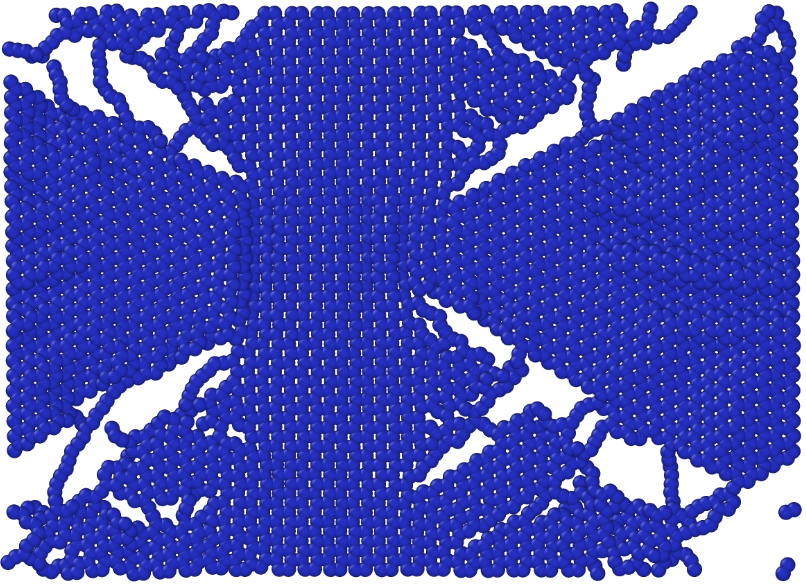}
        \caption{}
        \label{fig:uni_y}
    \end{subfigure}
    \hfill
    \begin{subfigure}[c]{0.32\linewidth}
        \centering
        \includegraphics[width=\linewidth]{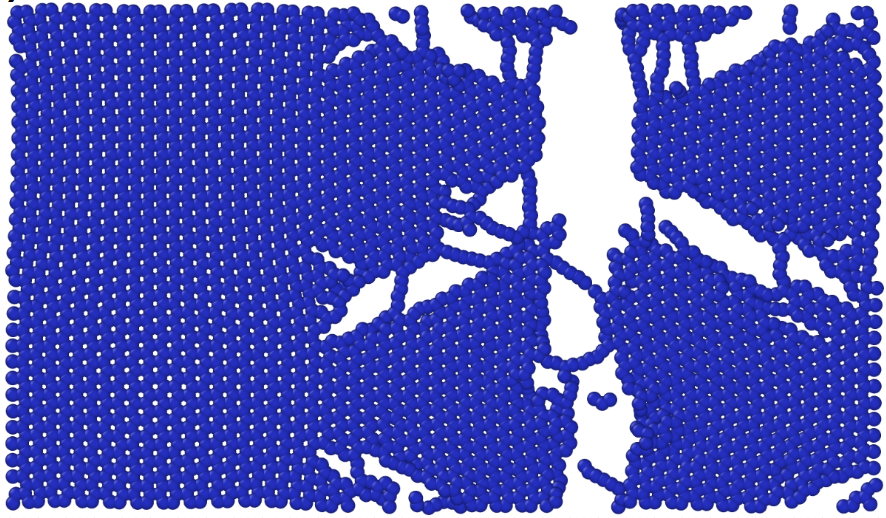}
        \caption{}
        \label{fig:bi}
    \end{subfigure}

    \caption{Visualization of the pristine graphene sheet at failure for (a) 0\textdegree{} uniaxial tension (armchair); (b) 90\textdegree{} uniaxial tension (zigzag); (c) pure biaxial tension.}
    \label{fig:sample_fracture}
\end{figure}

To establish a baseline for the nanoscopic fracture behavior of monocrystalline graphene, we first examine the defect-free scenario. This case represents the idealized, deterministic upper bound of the strength. It bears emphasis that simulations were performed at 273 K, and the quantitative results may be sensitive to this choice. Several representative fracture events for pristine graphene can be visualized in Figure~\ref{fig:sample_fracture}. 
The figure shows crack patterns for three loading scenarios: 0\textdegree{} (armchair) uniaxial tension, 90\textdegree{} (zigzag) uniaxial tension, and pure biaxial tension. These cases vividly illustrate the anisotropic nature of graphene’s lattice structure. Fracture tends to occur along the zigzag direction, even when the applied load acts counter to this path, revealing a strong crystallographic preference for zigzag-edge formation. In the biaxial case, cracks first nucleate along the vertical zigzag direction before branching diagonally, likely due to the shorter dimension of the sheet in the vertical direction, concentrating stress along that axis. This chirality dependence is consistent with prior simulation-based studies, as discussed by \citet{ni_anisotropic_2010}, \citet{zhang_fracture_2015}, and \citet{jiang_chirality-dependent_2017}.

\begin{figure}[ht!]
\centering
\includegraphics[width=0.55\linewidth]{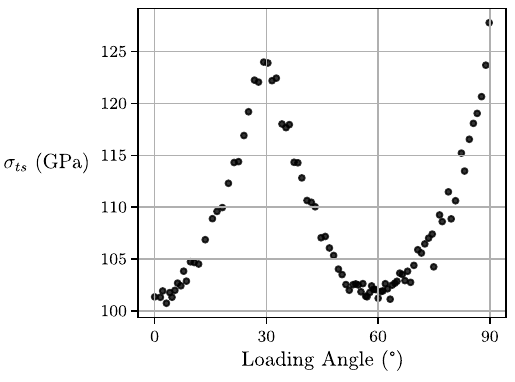}
\caption{\label{fig:aniso_uniaxial} Uniaxial tensile strength along varying loading angles, where 0\textdegree{} and 60\textdegree{} correspond to the armchair orientations, and 30\textdegree{} and 90\textdegree{} are the zigzag directions.}
\end{figure}

To further illustrate graphene’s anisotropic fracture response, Figure~\ref{fig:aniso_uniaxial} shows the uniaxial tensile strength plotted as a function of loading angle. 
The directional dependence is immediately apparent: the tensile strength peaks sharply near the two zigzag-aligned orientations at $\theta = 30$\textdegree{} and $\theta = 90$\textdegree, while reaching minima near the armchair-related directions. This angular variation reflects the intrinsic sixfold symmetry of the hexagonal lattice, which produces alternating strong and weak directions as the loading axis rotates. The periodic modulation of strength in this figure provides a clear depiction of the fundamental anisotropy that underlies graphene’s fracture behavior.

\subsubsection{Elastic Response of Pristine Graphene}
\begin{figure}[ht!]
\centering
\includegraphics[width=0.55\linewidth]{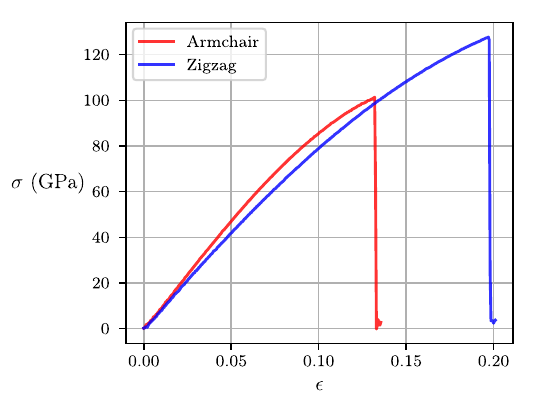}
\caption{\label{fig:sig_eps_pristine} Elastic response of pristine graphene in uniaxial tension for both the armchair and zigzag directions. Note the peak stress occurring directly before a sharp drop---this represents the uniaxial tensile strength along those directions.}
\end{figure}

Before examining strength as a whole, we first assess the initial elastic response of pristine graphene and validate it against findings reported in the literature. Figure~\ref{fig:sig_eps_pristine} shows the uniaxial stress–strain response of pristine graphene for loading aligned with $\theta = 0$\textdegree{} (armchair), shown in red, and $\theta = 90$\textdegree{} (zigzag), shown in blue. 
A slight directional dependence is observed in the elastic regime, reflecting the inherent anisotropy of the graphene lattice. From the linear portion of each curve, the Young’s modulus in the armchair direction is estimated as $E_{\text{AC}} \approx 950$~GPa, while in the zigzag direction it decreases to $E_{\text{ZZ}} \approx 830$~GPa. Both values, particularly the armchair direction, are consistent with the theoretically predicted in-plane stiffness of graphene, $E \approx 1$~TPa \citep{lee_measurement_2008}, confirming the accuracy of the simulation framework in reproducing the expected elastic behavior. 

\subsubsection{Deterministic Anisotropic Strength Surface Representation}
\begin{figure}[ht!]
\centering
\includegraphics[width=0.55\linewidth]{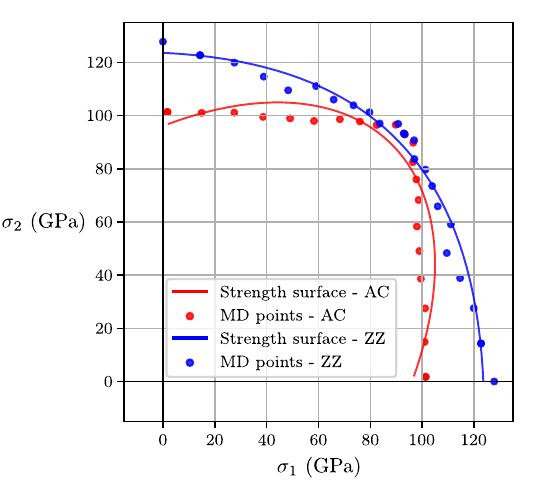}
\caption{\label{fig:dp_pristine_aniso} Strength surface and MD failure data of pristine graphene along armchair (AC) and zigzag (ZZ) directions.}
\end{figure}

As discussed, graphene exhibits a pronounced anisotropy in fracture strength at the single-crystal scale. Figure~\ref{fig:dp_pristine_aniso} compares MD failure points for loading in which the dominant principal direction is aligned with the armchair (red) and zigzag (blue) directions, along with their corresponding 2D strength surface model fits. 
Under uniaxial tension, the zigzag orientation attains a substantially higher fracture strength than the armchair orientation, and these points correspond to the $\theta=0$\textdegree{} and $\theta=90$\textdegree{} points in Figure~\ref{fig:aniso_uniaxial}. This difference diminishes as loading becomes more biaxial, finally becoming equivalent at perfect biaxial tension, where the loading angle is ambiguous. It is worth noting that the uniaxial tensile strength in the zigzag direction for graphene at this scale is observed to be 127.8 GPa. This is in remarkable agreement with the theoretical tensile strength, which is typically reported to be 130 GPa \citep{lee_measurement_2008}.

\begin{figure}[ht!]
    \centering
    \begin{subfigure}[t]{0.45\textwidth}
        \centering
        \includegraphics[width=\linewidth]{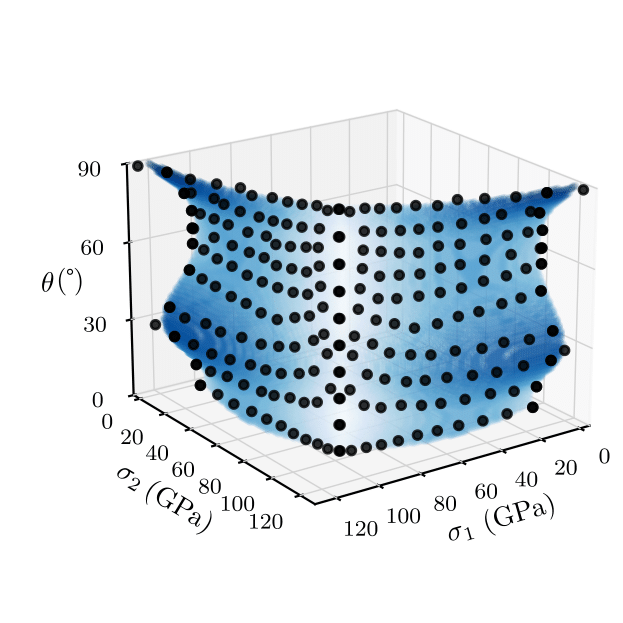}
        \caption{}
        \label{fig:3d_front}
    \end{subfigure}
    \hfill
    \begin{subfigure}[t]{0.45\textwidth}
        \centering
        \includegraphics[width=\linewidth]{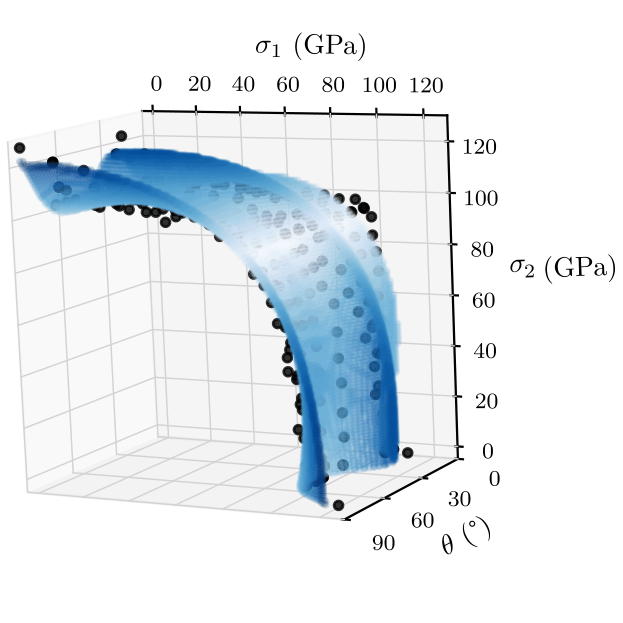}
        \caption{}
        \label{fig:3d_downview}
    \end{subfigure}
    \caption{Angular strength surface of pristine graphene (a) front view (b) top view. The black dots represent MD simulation results, while the blue shaded surface represents the fitted anisotropic strength surface. The color shading is solely for visualization purposes.}
    \label{fig:3D_pristine}
\end{figure}

With the ability to probe arbitrary loading directions and orientations relative to the lattice, we now deploy the framework to construct the three-dimensional orientation-dependent strength surface for pristine graphene. In this representation, $\sigma_1$ and $\sigma_2$ denote the principal stresses at failure, while the third axis $\theta$ corresponds to the orientation of the dominant principal direction with respect to the lattice. Figure~\ref{fig:3D_pristine} shows the MD failure points (black) together with the fitted angular strength surface model discussed in Section~\ref{sec:Anisotropic Strength Surface Modeling} (blue) from two different viewpoints. The surface closely follows the MD data across the full angular domain, capturing both the elevated uniaxial strength in the zigzag direction and the reduction toward the armchair direction, as well as the smooth variation between them. The color shading is included solely to aid visualization.

\subsection{\label{sec:Defective Graphene}Stochastic Anisotropic Strength Analysis of Defective Graphene}
We next extend our analysis to include the three types of defects mentioned in Section~\ref{sec:stochastic_angular_ss}, with the goal of quantifying the impact of added stochasticity in the fracture behavior. We will first look at the elastic response of defective graphene, aiming to understand the role defects play in the Young's modulus. Then, we will examine the building of the stochastic, anisotropic strength surface model, discussing modeling choices and overall performance, finally ending with conclusive insights. 

\subsubsection{\label{sec:elastic_response}Elastic Response of Defective Graphene}

\begin{figure}[ht!]
    \centering
    \begin{subfigure}[t]{0.48\textwidth}
        \centering
        \includegraphics[width=\linewidth]{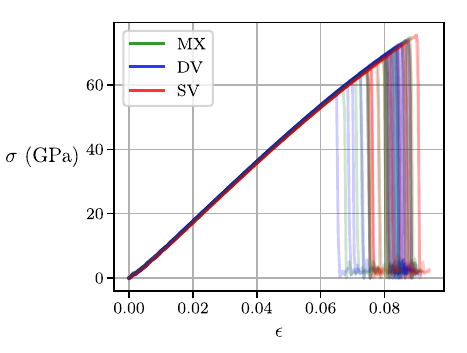}
        \caption{}
        \label{fig:sig_eps_ac}
    \end{subfigure}
    \hfill
    \begin{subfigure}[t]{0.48\textwidth}
        \centering
        \includegraphics[width=\linewidth]{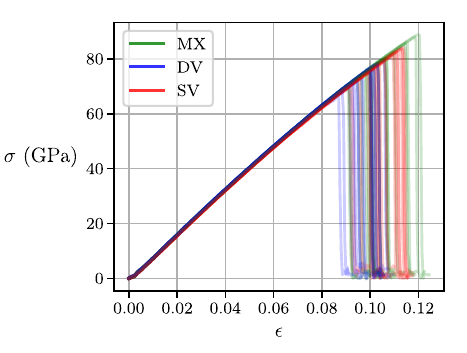}
        \caption{}
        \label{fig:sig_eps_zz}
    \end{subfigure}
    \caption{Elastic response of 20 realizations of defective graphene sheets for single vacancy (SV), double vacancy (DV) and mixed (MX) defect configurations, loaded uniaxially along the armchair (a) and zigzag (b) directions. While the initial slopes of the stress–strain curves exhibit minimal variability across realizations, the peak stresses display substantial scatter, highlighting the relative insensitivity of elasticity and the pronounced sensitivity of strength to defects.}
    \label{fig:sig_eps}
\end{figure}

Uniaxial tensile tests conducted on 20 independent defect realizations are displayed in Figure~\ref{fig:sig_eps}. 
These results show that the initial elastic response in both the armchair and zigzag directions is effectively identical across all three defect configurations. This confirms that the defect type does not influence the elastic modulus at small strains. It is worth noting, however, that the presence of defects slightly reduces the in-plane stiffness relative to pristine graphene, with $E_{\text{AC}_d} \approx 890$~GPa and $E_{\text{ZZ}_d} \approx 785$~GPa. This modest reduction reflects the loss of load-bearing bonds due to the atomic vacancies, but also suggests that the elastic response remains governed by the intact lattice network rather than by the local defect architecture. The relative similarity of the pristine and defective Young's moduli further indicates that the influence of defects manifests primarily in the fracture behavior, rather than in the initial elastic response of graphene.

\subsubsection{\label{sec:angular_ss_defective}Stochastic Anisotropic Strength Surface Representation}

We now extend the analysis from the armchair- and zigzag-aligned strength surfaces to the full angular domain, generating stochastic, orientation-dependent strength surfaces for each defect configuration. For each MD realization, the workflow described in Section~\ref{sec:Anisotropic Strength Surface Modeling} was applied to obtain the fitted $\theta$-dependent strength surface parameters $\alpha(\theta)$ and $k(\theta)$. This procedure was repeated across multiple random defect realizations for the three configurations---0.5\% SV, 0.5\% DV, and the mixed 0.25\% SV + 0.25\% DV case---yielding a distribution of angular strength surfaces for each.

\begin{figure}[ht!]
    \centering
    \begin{subfigure}[t]{0.48\textwidth}
        \centering
        \includegraphics[width=\linewidth]{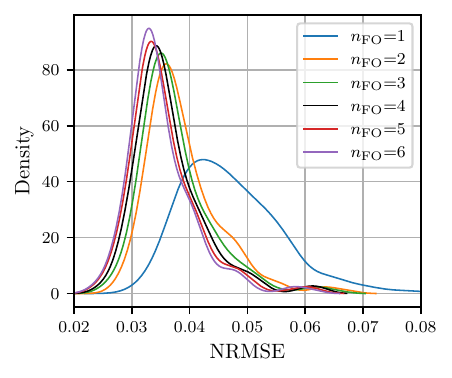}
        \caption{}
        \label{fig:nrmse}
    \end{subfigure}
    \hfill
    \begin{subfigure}[t]{0.48\textwidth}
        \centering
        \includegraphics[width=\linewidth]{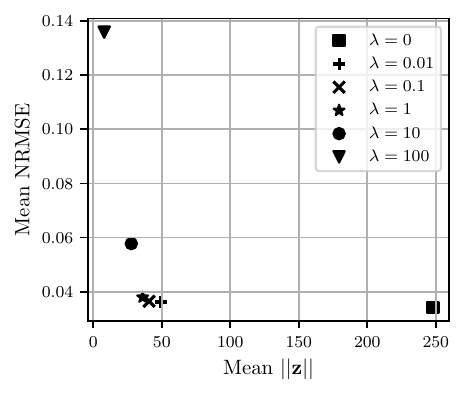}
        \caption{}
        \label{fig:lambda_conv}
    \end{subfigure}
    \caption{Evaluation of Fourier-order (a) and regularization choices (b) for fitting angular strength surface parameters. Both plots correspond to a concatenated dataset with all defect types present. (a) Convergence of the NRMSE as $n_{\text{FO}}$ increases for $\lambda=0.1$. Note that similar trends are observed for alternative values of $\lambda$. (b) Balance of the mean NRMSE and $||\mathbf z||$ for varying values of $\lambda$, where the Fourier order is fixed at $n_{\text{FO}}=4$. Note that similar trends are observed for alternative Fourier orders.}
    \label{fig:convergences}
\end{figure}

Fitting the strength surface parameter functions $\big(\alpha(\theta), k(\theta)\big)$ first requires determining the appropriate Fourier order, informed by NRMSE trends and qualitative inspection of the fitted angular functions. Figure~\ref{fig:nrmse} illustrates how the NRMSE distribution shifts as $n_{\text{FO}}$ increases. Reasonable convergence is observed as $n_{\text{FO}}$ increases, and the model begins to exhibit spurious oscillations for $n_{\text{FO}} \ge 5$. These nonphysical oscillations indicate overfitting arising from the data-driven fitting technique, particularly in regions where data are sparse. Based on these observations, in conjunction with the desire to decrease dimensionality, we select $n_{\text{FO}}=4$, corresponding to an 18-dimensional coefficient vector $\mathbf z$. Figure~\ref{fig:lambda_conv} shows the interaction of the NRMSE with $||\mathbf z||$ for various values of the regularization parameter $\lambda$. As described in Section~\ref{sec:Anisotropic Strength Surface Modeling}, the regularization term aims to control the magnitude of the $z_j$ coefficients without compromising the NRMSE. It is clear that $\lambda=0.1$ significantly reduces the magnitude of the coefficients without introducing any meaningful reduction of fit quality, so this was chosen for further analysis. 

\begin{figure}[ht!]
\centering
\includegraphics[width=0.9\linewidth]{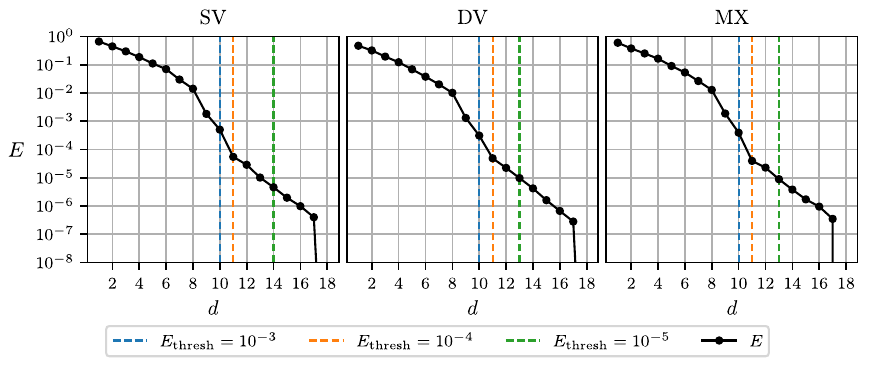}
\caption{\label{fig:pca} Decay of PCA recovery error $E$ as more principal components $d$ are retained for the single vacancy (SV), double vacancy (DV), and mixed (MX) defect datasets. Vertical dashed lines indicate representative recovery-error thresholds $E_{\text{thresh}}$, illustrating the trade-off between dimensionality reduction and fidelity of the reconstructed angular strength surfaces.}
\end{figure}

Figure~\ref{fig:pca} illustrates the decay of the recovery error $E(d)$ as additional principal components are retained. 
Due to the series of high-dimensional nonlinear transformations to map $\mathbf{z}$ to strength space, determining an appropriate recovery-error tolerance $E_{\text{thresh}}$ requires more than a purely statistical criterion. Our objective is to identify the smallest latent dimensionality for which the reconstructed $\mathbf{z}$ vectors still reproduce the essential features of the MD-derived strength surfaces.

\begin{figure}[ht!]
\centering
\includegraphics[width=0.8\linewidth]{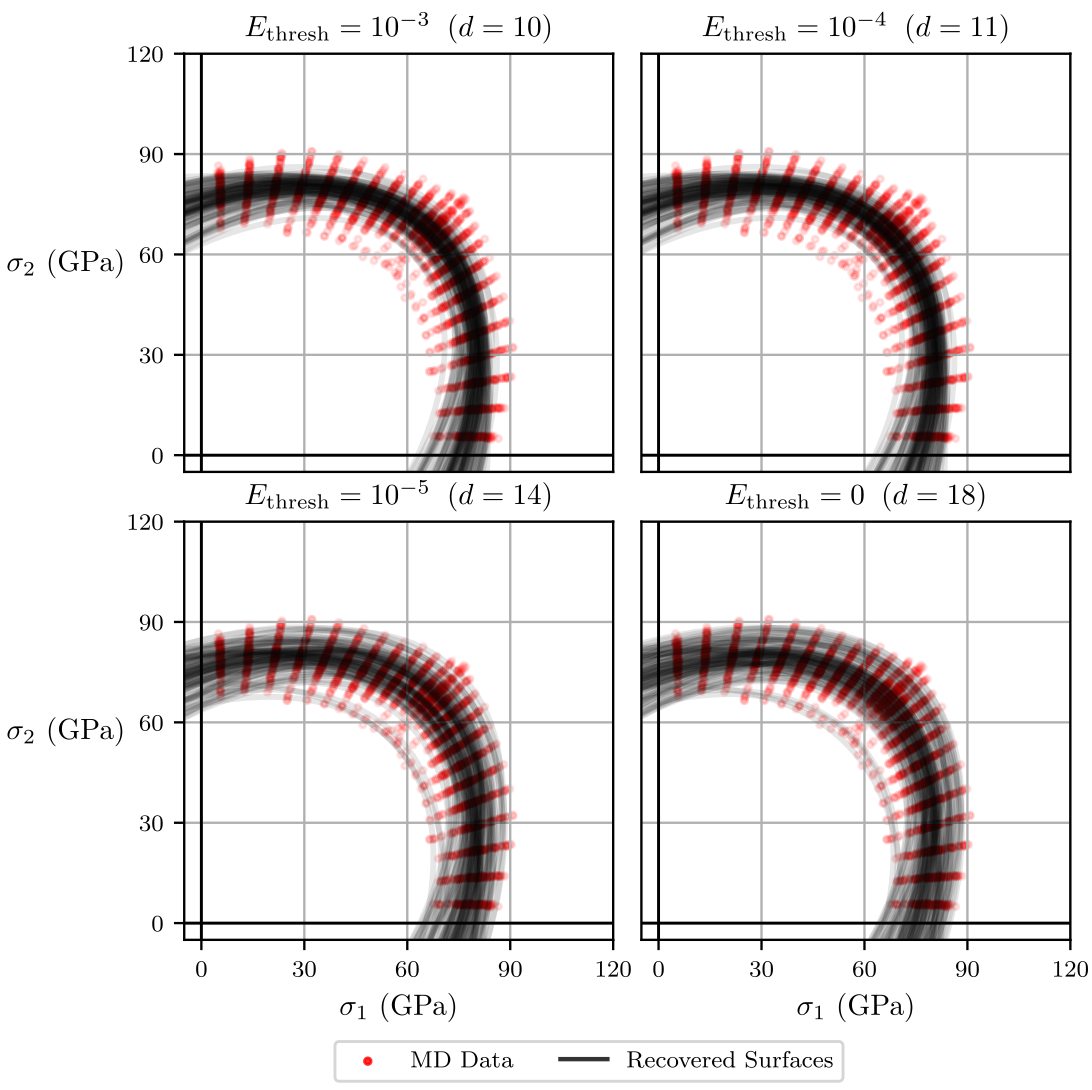}
\caption{\label{fig:pca_reconstruction} Visualization of reconstructed strength surfaces obtained by applying PCA dimensionality reduction to the original coefficient vectors $\mathbf z$, followed by inverse PCA to recover approximate coefficients. For each error threshold $E_{\text{thresh}}$, the reconstructed coefficients are evaluated at $\theta = 90$\textdegree{} and mapped into strength space, where they are compared directly against the original MD failure points. For simplicity, only the single-vacancy case at $\theta = 90$\textdegree{} is shown here, though the same trends persist for the double-vacancy and mixed-defect cases across other loading angles.}
\end{figure}

To assess this, Figure~\ref{fig:pca_reconstruction} compares reconstructed single vacancy $\mathbf{z}$ samples for several recovery error tolerances, evaluated at $\theta = 90$\textdegree. 
The case $E_{\text{thresh}}=0$ corresponds to the unreduced least-squares fit and serves as the reference against which all reduced representations are judged. The reconstructions indicate that a threshold of $E_{\text{thresh}} = 10^{-5}$ preserves the MD behavior with high fidelity. Relaxing the tolerance further leads to noticeable loss of structure in the reconstructed strength surfaces, reflecting the omission of physically relevant modes of variation. This analysis was repeated across all defect configurations and multiple angular slices, and in each case the same qualitative conclusion held: $E_{\text{thresh}} \approx 10^{-5}$ provides the lowest-dimensional latent representation that accurately captures the variability present in the MD dataset. Although this reduction aids in compressing the coefficient space, the resulting latent dimensionality remains relatively high, reflecting the inherent complexity of the angular strength surface and limiting the extent to which the representation can be simplified without sacrificing physical fidelity.

\begin{figure}[ht!]
    \centering
    \begin{subfigure}[c]{0.48\textwidth}
        \centering
        \includegraphics[width=\linewidth]{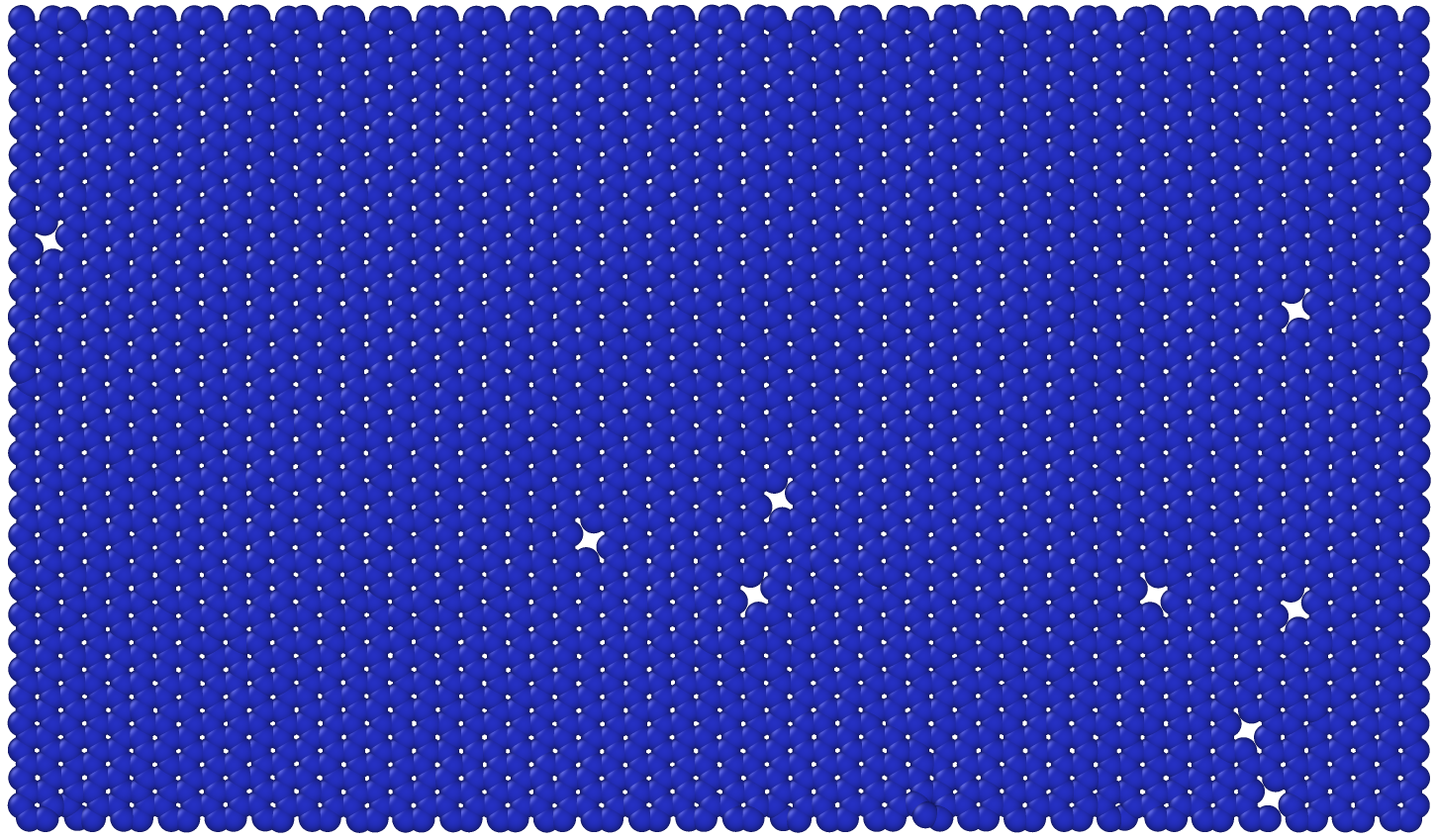}
        \caption{}
        \label{fig:outlier_vis}
    \end{subfigure}
    \hfill
    \begin{subfigure}[c]{0.48\textwidth}
        \centering
        \includegraphics[width=\linewidth]{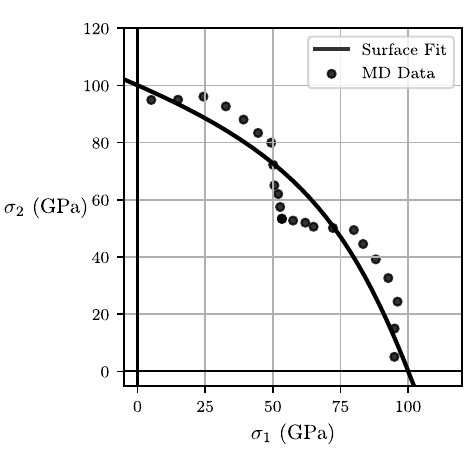}
        \caption{}
        \label{fig:outlier}
    \end{subfigure}
    \caption{(a) Visualization of a DV defect configuration sample that produces outlier star-shaped strength surface; (b) resulting jagged-like MD failure data and strength surface fit at $\theta=90$\textdegree.}
    \label{fig:outliers}
\end{figure}

After determining the proper amount of principal components to retain, we next select the number of Gaussian mixture modes based on the model’s ability to reproduce the distribution of the dataset in strength space. For the SV and MX cases, the latent-space statistics were well described by a single Gaussian component; no additional structure or clustering was observed, and a unimodal model was sufficient to capture the variability in their reconstructed strength surfaces. The DV case, however, exhibited notably different statistical behavior, as double vacancies are larger and less numerous, meaning they are more prone to forming spatial clusters. In several realizations, DV defects aggregated into elongated chains aligned approximately with the $y$–direction of the simulation cell, as shown in Figure~\ref{fig:outlier_vis}. 
This configuration is especially challenging: graphene is intrinsically weaker along the armchair ($x$) direction, and the defect alignment (facilitated in part by the shorter $y$–dimension of the sheet) locally amplifies this weakness. Under uniaxial loading at $\theta=90^\circ$, the sheet remains strong---often slightly stronger than typical realizations---but as the stress state becomes increasingly biaxial, load is transferred into the weakened $x$–direction, causing premature fracture along these DV-aligned paths. This produces a subset of strength surfaces that differ markedly from the majority of the dataset. Unlike the smooth, well-behaved envelopes observed in Figure~\ref{fig:pca_reconstruction}, these DV outliers exhibit jagged, star-shaped surfaces, clearly visible in the MD scatter shown in Figure~\ref{fig:outlier}. Importantly, these are not artifacts or noise; they are physically valid strength surfaces arising from realizations in which clustered DV defects dominated the fracture process. 

These atypical surfaces also expose a limitation of the two-parameter strength-surface formulation in Eq.~\eqref{eqn:DP}. Although this representation captures the overall trend---elevated uniaxial strength combined with notably reduced biaxial strength---its expressivity is too limited to replicate the fine irregularities of (pathological) star-shaped envelopes. The fit shown in Figure~\ref{fig:outlier} demonstrates this behavior: the general directional trend is preserved, but the detailed jagged structure is smoothed. 

\begin{figure}[ht!]
\centering
\includegraphics[width=0.8\linewidth]{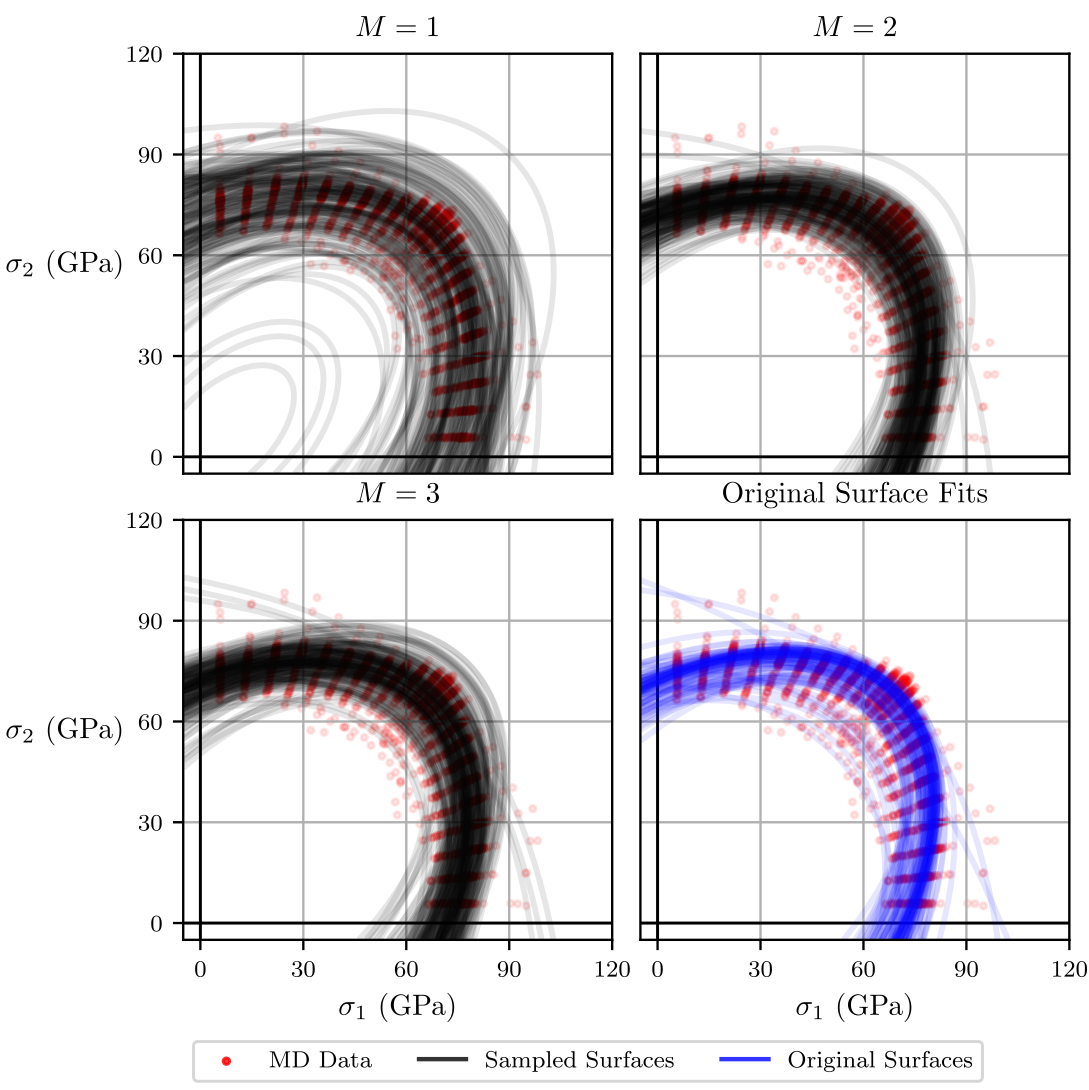}
\caption{\label{fig:gmm} Visualization of strength surfaces on top of the raw MD data for the DV dataset at $\theta=90$\textdegree. The three sets of black surfaces represent 200 instances of sampled strength surfaces from the latent space Gaussian mixture with the specified number of modes $M$. The blue surfaces represent the 101 original strength surface fits on top of the same MD data.}
\end{figure}

Because the underlying fracture behavior is physically meaningful, these realizations were retained in the dataset and incorporated into the statistical model. When projected into the latent PCA space, these atypical DV surfaces require additional Gaussian modes to reproduce. A single Gaussian component cannot capture both the dominant cluster of ``typical'' DV realizations and the smaller group of outlier, cluster-dominated strength surfaces. This was the main motivation behind the use of the Gaussian mixture model. Figure~\ref{fig:gmm} illustrates the resulting reconstructions at a fixed angle $\theta=90$\textdegree, with the original surface fits displayed in blue for reference: a unimodal model fails to reproduce the DV behavior altogether, a two-component model captures part of the structure but misses the extremes, whereas a three-component mixture successfully reconstructs both the conventional and star-shaped DV surfaces with an appropriate degree of variability. 

\subsubsection{Stochastic Anisotropic Strength Surface Analysis}\label{sec:Stochastic Anisotropic Strength Surface Analysis}

\begin{figure}[ht!]
    \centering
    \begin{subfigure}[c]{0.85\linewidth}
        \centering
        \includegraphics[width=\linewidth]{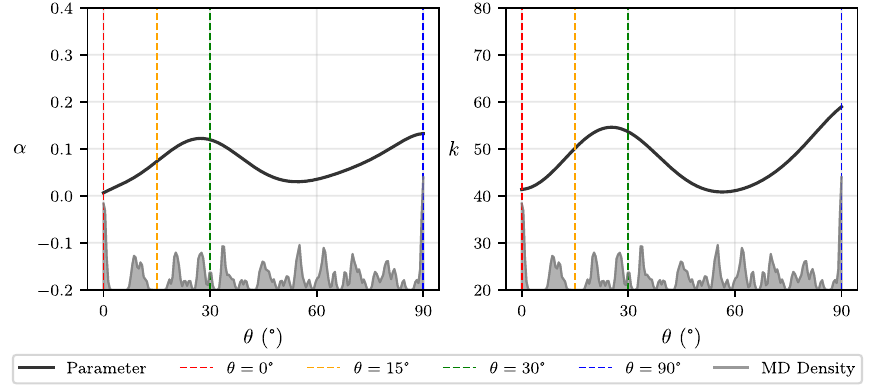}
        \caption{Instance of fitted $\alpha(\theta)$ and $k(\theta)$ (black) and density of sample's MD data (gray).}
        \label{fig:ak_functions}
    \end{subfigure}

    \vspace{1em}
    \begin{subfigure}[c]{0.75\linewidth}
        \centering
        \includegraphics[width=\linewidth]{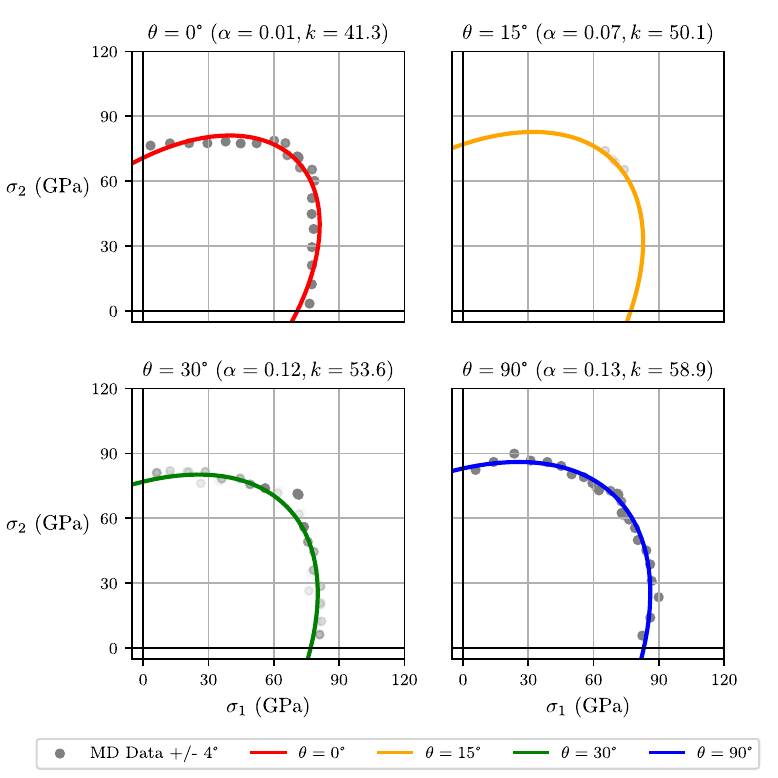}
        \caption{Sliced plane stress strength surfaces at $\theta=[0, 15, 30, 90]$\textdegree. The opacity of the gray MD datapoints represent the proximity to the desired angle ($\pm4$\textdegree).}
        \label{fig:sample_dps}
    \end{subfigure}

    \caption{Full continuous parameter functions and sampled slices of strength space. Each color in (a) represents a slice of the strength surface parameters at a certain angle, and (b) shows the constructed surface at that slice.}
    \label{fig:ak_and_sample_dps}
\end{figure}

With the ability to fit, model, and sample new anisotropic strength surfaces, we finally turn to interpreting the resulting angular parameters and what they reveal about graphene’s fracture behavior at this scale. To understand the fitted angular functions $\big(\alpha(\theta), k(\theta)\big)$ and their implications for strength, Figure~\ref{fig:ak_and_sample_dps} presents a representative example in which the parameter curves appear in Figure~\ref{fig:ak_functions} and the corresponding strength-surface slices appear in Figure~\ref{fig:sample_dps}. Although the plotted $\big(\alpha(\theta), k(\theta)\big)$ functions may visually resemble a sixfold symmetric pattern, this is specific to the particular defect realization shown; vacancy distributions generally break lattice symmetry, and the fitted functions do not enforce any such repeating structure. By evaluating $\alpha(\theta)$ and $k(\theta)$ at arbitrary angles, we can reconstruct the corresponding two-dimensional strength surface via Eq.~\eqref{eqn:DP}, providing a direct depiction of how the failure envelope evolves with orientation. The example also illustrates the strong correlation between $\alpha$ and $k$, reflecting how these parameters jointly govern the curvature and location of the strength boundary.

The gray distribution beneath the curves in Figure~\ref{fig:ak_functions} denotes the density of MD failure data as a function of $\theta$ for this specific defect realization, highlighting regions of abundant sampling as well as angles where data are sparse. This sampling density is also reflected in the opacity of the MD points in Figure~\ref{fig:sample_dps}: only points within 4\textdegree{} of the target angle for a given slice are shown, with opacity varying smoothly from fully opaque when the point matches the target angle to fully transparent as the angular mismatch approaches 4\textdegree. Together, these visualizations underscore a key advantage of the angular formulation in Eq.~\eqref{eqn:angular_dp}: it provides a continuous, physically admissible representation of the strength surface even in angular regions with limited MD coverage. In this instance, although few MD data points exist near $\theta = 15$\textdegree, the $\big(\alpha(\theta), k(\theta)\big)$ functions interpolate smoothly through this interval, filling in gaps between sampled orientations while remaining consistent with the broader physical trends present in the MD simulations.

\begin{figure}[ht!]
\centering
\includegraphics[width=0.9\linewidth]{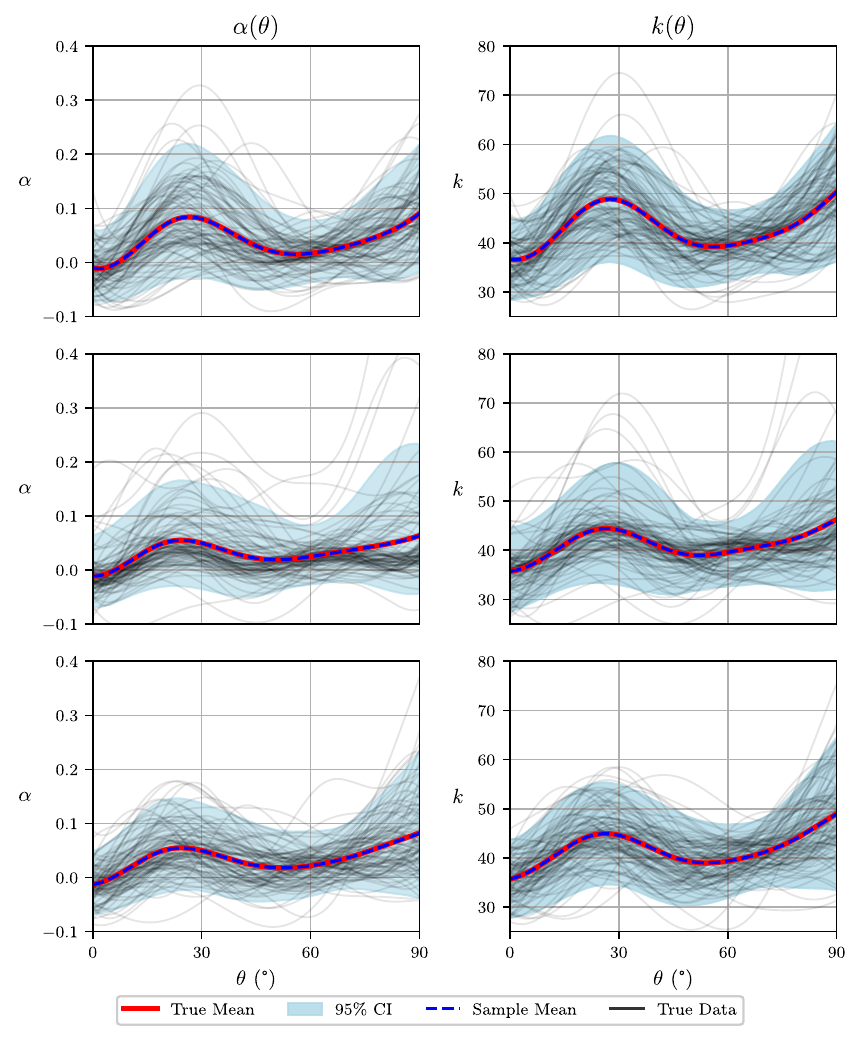}
\caption{\label{fig:param_ci} Visualization of the sampled mean and confidence intervals on top of the original dataset for $\alpha$ (left) and $k$ (right). The top row represents the single vacancy case, the middle row represents the double vacancy case, and the bottom row represents the mix of single and double vacancies.}
\end{figure}

\clearpage
The fitted angular parameters $\alpha(\theta)$ and $k(\theta)$ obtained from each random defect realization across all three defect configurations are shown as faint black curves in Figure~\ref{fig:param_ci}. 
The light-blue envelopes  denote the 95\% confidence intervals derived from the sampled functions $\hat \alpha(\theta)$ and $\hat k(\theta)$. The close agreement between the mean of the sampled parameters and the mean of the original dataset, together with the ability of the 95\% confidence intervals to encompass nearly but not quite all of the original data, validates the stochastic model and demonstrates that the auxiliary transformation-PCA-Gaussian mixture framework can accurately reproduce the underlying molecular-scale variability. It is worth highlighting the few parameter curves in the DV case that experience a large spike at $\theta=90$\textdegree, traveling out of the frame of reference. These correspond with the samples where we see a more jagged star-shaped structure. 

Another notable feature in Figure~\ref{fig:param_ci} is the behavior of angular periodicity in the presence of defects. At the level of individual realizations, the fitted parameter functions no longer exhibit the sixfold symmetry associated with the pristine lattice, reflecting the symmetry-breaking effect of random defect distributions. However, when considered collectively through the stochastic model, a residual, approximate sixfold structure emerges in the mean response and confidence envelopes. This behavior indicates that while defects destroy exact rotational symmetry at the realization level, the underlying lattice symmetry remains statistically imprinted in the aggregate strength response.

\begin{figure}[ht!]
\centering
\includegraphics[width=0.9\linewidth]{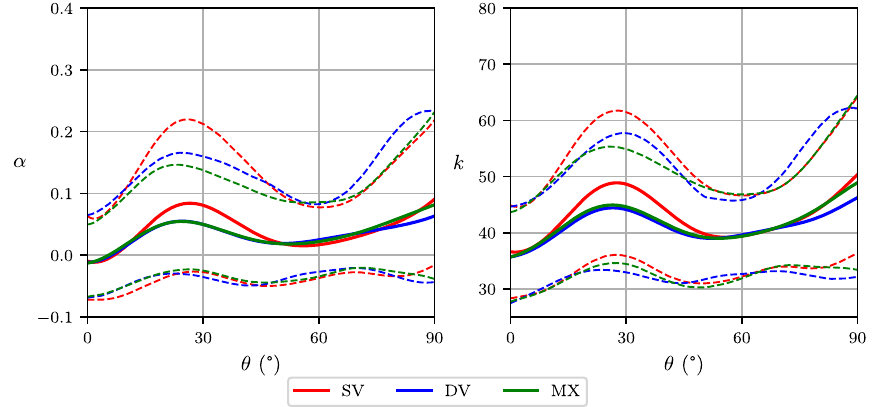}
\caption{\label{fig:all_cis} Comparison of means (solid line) and 95\% confidence intervals (dashed line) of strength surface parameters ($\alpha$ left, $k$ right) for single vacancy (SV), double vacancy (DV) and mixed (MX) defect types.}
\end{figure}

Figure~\ref{fig:all_cis} compares the mean and 95\% confidence intervals for $\alpha(\theta)$ and $k(\theta)$ across all three defect configurations on shared axes. 
The strong overlap among the curves demonstrates that, within the examined defect density, the type of defect---single vacancy, double vacancy, or mixed---has a negligible effect on the anisotropic fracture response. Instead, the dominant factor governing variability in the angular strength surface is the overall defect density. 

\begin{figure}[ht!]
\centering
\includegraphics[width=0.9\linewidth]{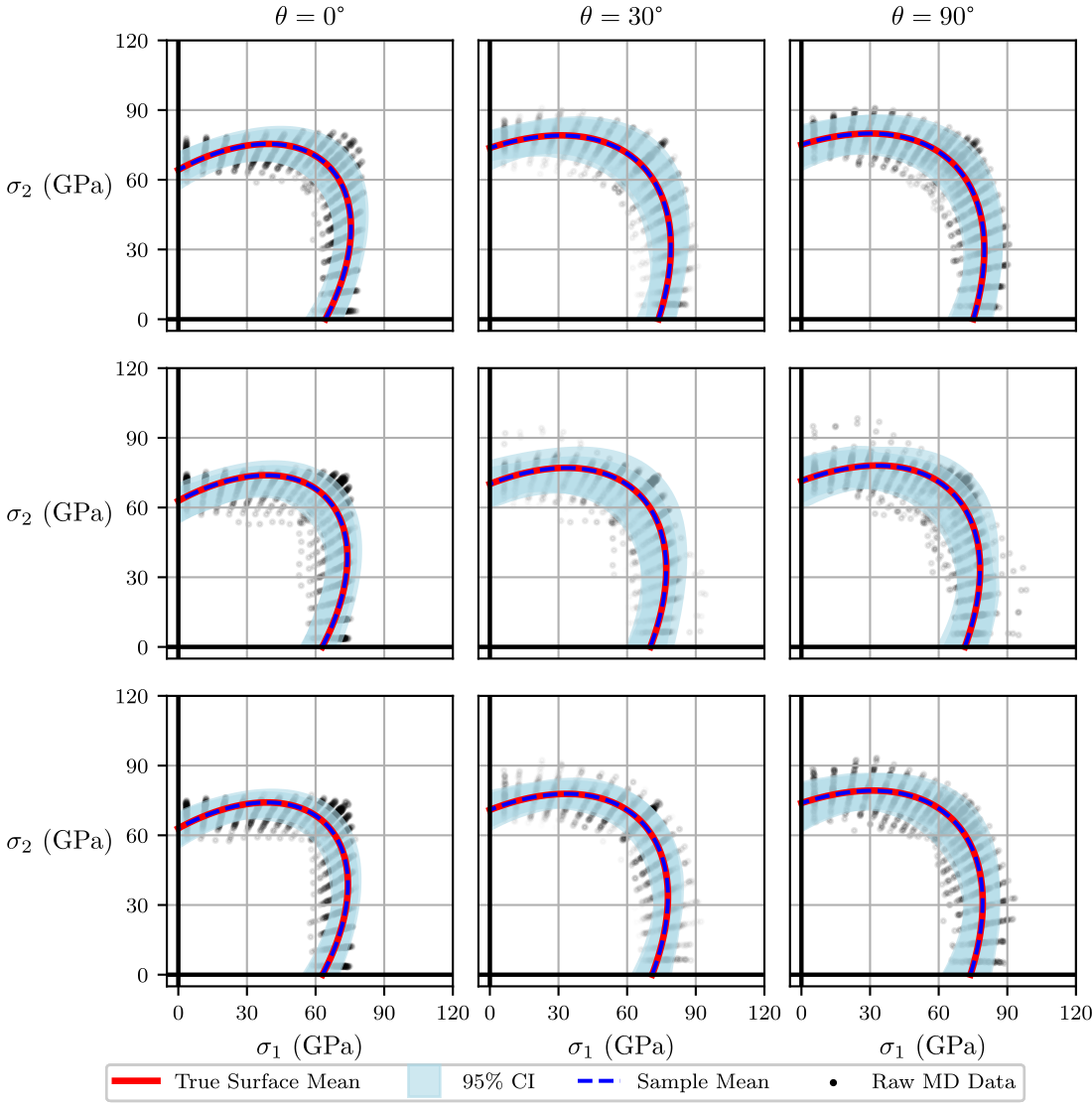}
\caption{\label{fig:strength_ci} Visualization of the sampled mean and confidence intervals directly in stress space, at angles of interest $\theta=[0, 30, 90]$\textdegree. The top row represents the single vacancy case, the middle row represents the double vacancy case, and the bottom row represents the mix of single and double vacancies. The statistical model is not calibrated to enforce exact reproduction of the molecular dynamics data at every angle; rather, the intervals reflect variability inferred from the sampled angular strength surfaces.}
\end{figure}

Finally, by sampling new realizations of $\big(\alpha(\theta), k(\theta)\big)$ and mapping them back into stress space, we obtain mean curves and confidence intervals for the resulting strength surfaces. Figure~\ref{fig:strength_ci} illustrates these intervals at $\theta = [0, 30, 90]$\textdegree{} for the SV, DV, and MX defect cases. 
The close agreement among the three configurations again confirms that defect type plays only a minor role in shaping graphene's general strength; the fracture response is governed instead by how many atoms are missing, not which type of vacancy created the missing atoms.

Taken together, these findings reinforce that, for sufficiently dilute vacancy populations, defective graphene can be statistically described using a unified defect-density framework, largely independent of defect morphology. The main exception arises as domain size shrinks, where double vacancies may occasionally cluster and induce localized anisotropic weakening. Although this effect leaves only a subtle imprint on the confidence intervals, it appears more clearly in the latent-space sampling behavior for the DV case, and is worth accounting for in models. Thus, while defect morphology becomes negligible in large domains where clustering is unlikely, it may still play a nontrivial role in smaller systems where defect interactions are more probable.

\section{Conclusion}

We have presented a framework for quantifying the stochastic, anisotropic fracture behavior of materials from the perspective of strength, demonstrated here using monocrystalline graphene through molecular dynamics simulations and statistical inference. The methodology utilizes a data-driven strain-stress mapping to produce an orientation-dependent strength surface representation, then applies transformation and probabilistic modeling techniques to capture both the directional dependence and the variability introduced by atomic-scale defects.

Applied to pristine graphene, the proposed angular strength surface model accurately represented molecular dynamics failure data across all loading orientations, capturing the elevated tensile strength in the zigzag direction and the reduced strength in the armchair direction, consistent with the known sixfold anisotropy of the lattice. Introducing single and double vacancy defects at fixed densities added stochastic variability, slightly altered the overall elastic response, and disrupted the sixfold symmetry of individual strength surfaces. The defects resulted in reduced strengths of graphene overall, as well as a dispersion in the strengths, particularly at orientations aligning with the zigzag direction. 

By fitting each defect realization with a smooth, physically constrained angular strength surface, the fracture behavior was reduced to a compact vector of Fourier coefficients. Given the substantial computational effort required to generate each realization, Principal Component Analysis was employed to extract a low-dimensional latent representation from a limited ensemble of strength surfaces, which was subsequently well characterized by a multivariate Gaussian mixture model. The Gaussian mixture model was necessary to capture the clustering tendencies of larger double vacancy defects, where jagged, star-shaped outlier surfaces were more likely to arise. The ability for the Gaussian mixture model to collapse back to a singular multivariate Gaussian allowed for convenient modeling in situations where these outliers did not exist. Sampling from this probabilistic model enabled the generation of new, statistically consistent anisotropic strength surfaces. These could be used to construct confidence intervals in the parameter space, as well as directly in stress space, for arbitrary loading angles.

These confidence intervals provide a direct quantitative measure of uncertainty in graphene’s anisotropic strength, effectively bounding the range of admissible fracture strengths for any given loading direction and defect configuration. In doing so, the angular strength surface transitions from a purely descriptive construct into a predictive statistical tool, one capable of informing uncertainty-aware material models and guiding engineering-scale design decisions. Although demonstrated on graphene using a particular interatomic potential and a specific parametric strength surface, the proposed framework is intentionally agnostic to material system, choice of interatomic potential, and strength surface formulation, inheriting all constitutive behavior directly from the underlying MD description. It may therefore be extended directly to other materials, potentials, defect architectures, or alternative admissible strength models without modification to the underlying inference procedure. It also represents a generalizing step toward three-dimensional or polycrystalline systems, thereby bridging atomistic simulations with continuum-scale modeling through a unified, data-driven methodology. More broadly, the statistical insight gained at the monocrystalline scale provides essential input for potential mesoscale polycrystalline models, supporting a continued bottom-up effort to characterize strength and ultimately improve the predictive capabilities of fracture simulations.

\section*{Code Availability}
All scripts used to set up simulations, including datafile generation, LAMMPS input files, automation and data storage scripts, as well as all analysis and post-processing codes used to produce the results presented in this work, are publicly available at \url{https://github.com/abon11/Graphene_Strength_Surface}.

\appendix

\section{\label{app:nn_mapping}Nonlinear Strain-Stress Mapping}
Because the stress response of graphene is a nonlinear function of both loading ratio and orientation, a purely analytical mapping from imposed strain-rate tensors to resulting stress tensors is not feasible. We therefore developed a data-driven surrogate model to approximate this mapping, enabling us to prescribe arbitrary combinations of stress ratio and orientation.

A dataset was generated by applying randomly oriented and proportioned in-plane strain-rate tensors to pristine graphene and recording the steady-state stress tensors after a fixed simulation time. The normal strain-rate components were constrained to be nonnegative so that only tensile loads were applied, and the shear component was constricted to the positive direction to focus the angle of loading in $\theta \in [0, 90]$\textdegree. The maximum strain rate in any direction was $\dot \varepsilon_{\max}$. If some loading configuration led to buckling (e.g., pure shear), the simulation was terminated early to ensure that the dataset contained only valid tensile states. In total, 38,123 simulations were performed. To ensure adequate stress development and avoid under-loaded configurations, simulations with $\sigma_1 \le 4~\mathrm{GPa}$ were discarded, yielding a final dataset of 34,142 simulations.

To learn the mapping, we trained a multi-output feedforward neural network (MLP) composed of three hidden layers with 256, 128, and 64 neurons, respectively. The ReLU activation function was used for all layers. This architecture suffices to obtain reasonable accuracy for the applications considered in this paper, and no attempt was made to further optimize the network model. Optimization was performed using the Adam algorithm, with early stopping. Inputs ($\dot \varepsilon_x, \dot \varepsilon_y, \dot \varepsilon_{xy}$) and outputs ($\sigma_x, \sigma_y, \sigma_{xy}$) were standardized to zero mean and unit variance before training. Hyperparameter tuning was achieved considering an 80/20 train-test split, with both inputs and outputs unscaled after prediction. Although other regression approaches were tested (gradient-boosted trees, symbolic regression), the MLP provided the best balance of accuracy and computational efficiency.

To validate the fit of the neural network surrogate, we report both raw and normalized error metrics. Let $\sigma_k$ denote a stress component and $\widehat{\sigma}_k$ its prediction; the normalized mean squared error for component $k$ is defined as
\begin{equation}
    \text{nMSE}_k = \frac{\text{MSE}(\sigma_k, \widehat \sigma_k)}{\max(\sigma_k^{\text{test}})^2},
\end{equation}
where $\max(\sigma_k^\text{test})$ is the maximum magnitude of component $k$ in the test set. This normalization places errors for all stress components on a common scale, accounting for the smaller dynamic range of $\sigma_{xy}$ relative to the normal stresses. On the held-out test data, the trained model achieved nMSE values of $2.47 \times 10^{-4}$, $1.96 \times 10^{-4}$, and $9.21 \times 10^{-4}$ for $\sigma_x$, $\sigma_y$, and $\sigma_{xy}$ respectively. The close agreement between the final overall training and test raw MSE values (0.021515 vs.\ 0.021578) confirms that the model did not overfit. The corresponding coefficients of determination were $R^2_{\sigma_x} = 0.996$, $R^2_{\sigma_y} = 0.997$, and $R^2_{\sigma_{xy}} = 0.986$, indicating that over 98\% of the variance in the MD stress data is captured by the model. 

\begin{figure}[ht!]
    \centering

    \includegraphics[width=0.8\linewidth]{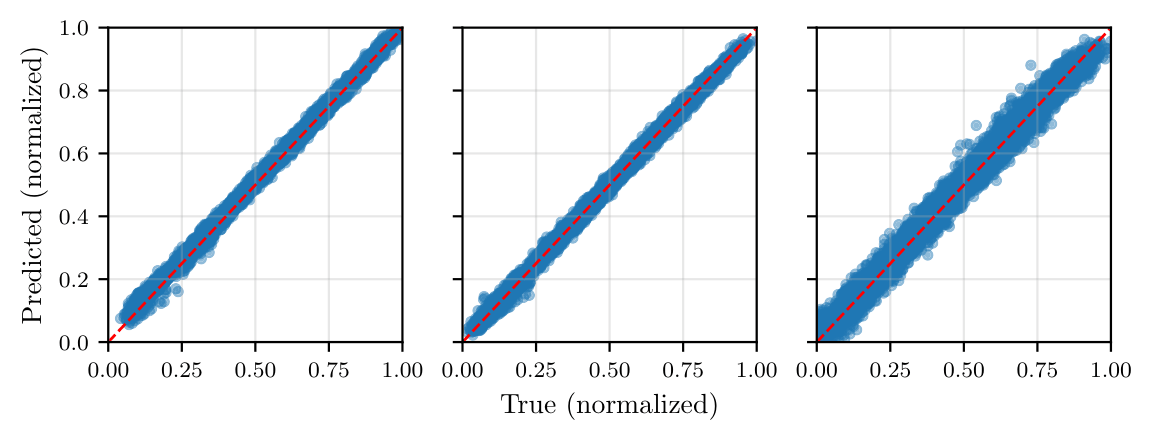}

    \caption{Predicted versus true stress components for the neural network strain–stress surrogate, evaluated on the held-out test set for $\sigma_x$, $\sigma_y$, and $\sigma_{xy}$ (left to right). All quantities are normalized by the maximum magnitude of the corresponding stress component to enable direct comparison across components. The dashed line indicates perfect agreement. The close alignment of the predictions with the reference values across the full range confirms the accuracy of the model, with slightly increased scatter observed for $\sigma_{xy}$ due to the greater nonlinearity of the shear response.}
    \label{fig:pva}
\end{figure}

This performance is visualized in the predicted–versus–true comparison shown in Figure~\ref{fig:pva}, where the predicted stresses closely follow the MD reference values across all components. The slight degradation in accuracy for $\sigma_{xy}$ is expected, as the shear response is more nonlinear and sensitive to local atomic rearrangements. Nevertheless, the overall agreement confirms that the mapping from strain rate to stress is sufficiently accurate for use in the inverse formulation described below.

This model was used to target specific loading states, solving an inverse problem. The methodology proceeds as follows. We first define desired stress ratio and orientation, and then convert this information into a Cartesian stress tensor via rotation of the principal stresses. We subsequently solve a bound-constrained optimization problem to identify a strain-rate tensor that minimizes the squared error between the predicted stresses and the target stresses. This optimization is performed in scaled space, for numerical stability, and subject to nonnegativity and magnitude bounds on the strain rates---to avoid compressive buckling and ensure moderate strain rate, $\dot \varepsilon_{\max} \leq 0.001$. Differential evolution is used for the three-variable inverse search, embedded within an outer scalar loop on the magnitude of the principal stress. In practice, we iteratively request a target stress tensor corresponding to a trial value of $\sigma_1$, solve the inverse problem to obtain the strain-rate tensor that most closely reproduces this stress state, and then check whether the maximum component of the recovered strain-rate tensor exceeds the prescribed cap $\dot \varepsilon_{\max}$. If it does, the trial $\sigma_1$ is adjusted, using Brent’s method root-finding routine. This iterative process continues until convergence to a consistent pair---a target stress tensor and its inverse-mapped strain-rate tensor required to lie entirely within the admissible bounds and have its largest entry approximately equal to $\dot \varepsilon_{\max}$---is obtained. The recovered strain-rate tensors then serve as the loading inputs for MD simulations at arbitrary ratio–orientation combinations.

The ability of the resulting model to generate requested principal stress ratios at prescribed loading angles was validated and found to be sufficiently accurate for the inverse loading formulation employed in this work. Some limitations were nevertheless observed. In particular, the model struggles to reproduce prescribed loading angles in the limit case $\sigma_2/\sigma_1 \rightarrow 1$, where the distinction between principal directions becomes progressively weaker, corresponding to an approach toward a physically ill-defined loading orientation. In addition, the model exhibits some difficulty in producing uniaxial tension along or near $\theta = 90^\circ$, owing to the absence of compressive loading states in the training data. This limitation was addressed using the procedures described in Section~\ref{sec:Orientation-Fixed Strength Surface Construction}.

\section{Fracture Detection Algorithm}
\label{app:fracture_alg}
Algorithm~\ref{alg:fracture_gated} summarizes the fracture-detection procedure used to identify MD failure events from noisy principal stress histories---both on the fly to trigger termination of the simulation and upon conclusion of the simulation to extract strength and fracture time results.

\begin{algorithm}[H]
\caption{Fracture detection}
\label{alg:fracture_gated}
\begin{algorithmic}[1]
\Require Principal stress history $\{\sigma_1(t),\sigma_2(t)\}$ sampled every $\Delta t$; check window sizes $n_{\text{last}}$, $n_{\text{prev}}$; degradation factor $\eta\in(0,1)$; drop threshold $\Delta\sigma$
\Ensure Strength $\boldsymbol{\sigma}^\ast$, fracture time $t_f$ \Comment{Return None if not all checks pass}

\State $\boldsymbol{\sigma}^\ast \gets \text{None}$, $t_f \gets \text{None}$
\State $N \gets \text{len}\{\sigma_1(t),\sigma_2(t)\}$
\State \textbf{Check 1: History length} 
\If{$N < n_{\text{prev}} + n_{\text{last}}$} \Return no fracture \Comment{Ensure sufficient sampling history} \EndIf

\State \textbf{Check 2: Running mean drop}
\For{$i \in \{1,2\}$}
    \State $\overline{\sigma}_{i,\text{last}} \gets \frac{1}{n_{\text{last}}}\sum_{k=N-n_{\text{last}}+1}^{N} \sigma_i(\Delta t \cdot k)$ \Comment{Avg stress for last $n_{\text{last}}$ time steps}
    \State $\overline{\sigma}_{i,\text{prev}} \gets \frac{1}{n_{\text{prev}}}\sum_{k=N-n_{\text{last}}-n_{\text{prev}}+1}^{N-n_{\text{last}}} \sigma_i(\Delta t \cdot k)$ \Comment{Avg stress for prev $n_{\text{prev}}$ time steps before that}
    \State $\text{drop}_i \gets \big(\overline{\sigma}_{i,\text{last}} < \eta\,\overline{\sigma}_{i,\text{prev}}\big)$ \Comment{Bool if there is drop in avg stress}
\EndFor
\If{not $(\text{drop}_1 \lor \text{drop}_2)$} \Return no fracture \Comment{Return None if no drops detected}\EndIf
\State $I \gets \{\, i \in \{1,2\} \mid \text{drop}_i = \text{true} \,\}$ \Comment{Store indices where drop is detected}

\State \textbf{Check 3: Peak existence and location}
\For{$i \in I$} 
    \State Find peaks $\mathcal{P}_i = \{t_p\}$ in $\sigma_i(t)$
    \If{$\mathcal{P}_i = \emptyset$} \Return no fracture \EndIf
    \State $t_{p}^{(i)} \gets \arg\max_{t\in \mathcal{P}_i}\, \sigma_i(t)$ \Comment{Find largest peak index}
    \If{$t_{p}^{(i)} \ge N - n_{\text{last}} + 1$}
        \State $s^{\text{low}}_i \gets \min\{\sigma_i(t)\,|\, t \in [N-n_{\text{last}}+1,\,N]\}$ \Comment{Look for min stress at end of simulation}
    \Else
        \State $s^{\text{low}}_i \gets \min\{\sigma_i(t)\,|\, t \in [t_{p,i}+1,\, t_{p,i}+n_{\text{last}}]\}$ \Comment{Look for min stress in region directly after peak}
    \EndIf
    \State \textbf{Check 4: Drop magnitude} 
    \State $\text{dropOK}_i \gets (\sigma_i(t_{p}^{(i)}) - s^{\text{low}}_i \ge \Delta\sigma)$ \Comment{Check if stress drop is large enough}
    \If{$\text{dropOK}_i$}
        \State \texttt{fracture\_found} $\gets$ \texttt{true} \Comment{Mark fracture detected and stop checking secondary indices}
        \State \textbf{break}
    \Else
        \State \texttt{fracture\_found} $\gets$ \texttt{false} \Comment{Continue to next index if any}
    \EndIf
\EndFor

\If{not \texttt{fracture\_found}} \Return no fracture \EndIf
\State \textbf{Declare fracture and record outputs}
\State Prioritize $i^\ast = 1$ \Comment{Break ties by defaulting to larger stress value}
\State $\boldsymbol{\sigma}^\ast \gets \big(\sigma_1(t_{p}^{(i^\ast)}),\,\sigma_2(t_{p}^{(i^\ast)}),\,\sigma_3(t_{p}^{(i^\ast)})\big)$
\State $t_f \gets t_{p}^{(i^\ast)}\,\Delta t$
\State \Return $\boldsymbol{\sigma}^\ast,\,t_f$ \Comment{Return strength and fracture time}
\end{algorithmic}
\end{algorithm}

\bibliographystyle{elsarticle-harv}
\bibliography{references}

@article{kumar_phase-field_2020,
	title = {The phase-field approach to self-healable fracture of elastomers: {A} model accounting for fracture nucleation at large, with application to a class of conspicuous experiments},
	volume = {107},
	issn = {01678442},
	shorttitle = {The phase-field approach to self-healable fracture of elastomers},
	url = {https://linkinghub.elsevier.com/retrieve/pii/S0167844219307566},
	doi = {10.1016/j.tafmec.2020.102550},
	language = {en},
	urldate = {2025-11-05},
	journal = {Theoretical and Applied Fracture Mechanics},
	author = {Kumar, Aditya and Lopez-Pamies, Oscar},
	month = jun,
	year = {2020},
	pages = {102550},
}

@article{kumar_revisiting_2020,
	title = {Revisiting nucleation in the phase-field approach to brittle fracture},
	volume = {142},
	issn = {00225096},
	url = {https://linkinghub.elsevier.com/retrieve/pii/S0022509620302623},
	doi = {10.1016/j.jmps.2020.104027},
	language = {en},
	urldate = {2025-11-05},
	journal = {Journal of the Mechanics and Physics of Solids},
	author = {Kumar, Aditya and Bourdin, Blaise and Francfort, Gilles A. and Lopez-Pamies, Oscar},
	month = sep,
	year = {2020},
	pages = {104027},
}

@article{jiang_chirality-dependent_2017,
	title = {The chirality-dependent fracture properties of single-layer graphene sheets: {Molecular} dynamics simulations and finite element method},
	volume = {122},
	issn = {0021-8979},
	shorttitle = {The chirality-dependent fracture properties of single-layer graphene sheets},
	url = {https://doi.org/10.1063/1.4993176},
	doi = {10.1063/1.4993176},
	number = {2},
	urldate = {2024-11-05},
	journal = {Journal of Applied Physics},
	author = {Jiang, Zonghuiyi and Lin, Rong and Yu, Peishi and Liu, Yu and Wei, Ning and Zhao, Junhua},
	month = jul,
	year = {2017},
	pages = {025110},
}

@article{ni_anisotropic_2010,
	title = {Anisotropic mechanical properties of graphene sheets from molecular dynamics},
	volume = {405},
	issn = {0921-4526},
	url = {https://www.sciencedirect.com/science/article/pii/S0921452609014379},
	doi = {10.1016/j.physb.2009.11.071},
	number = {5},
	urldate = {2024-12-03},
	journal = {Physica B: Condensed Matter},
	author = {Ni, Zhonghua and Bu, Hao and Zou, Min and Yi, Hong and Bi, Kedong and Chen, Yunfei},
	month = mar,
	year = {2010},
	pages = {1301--1306},
}

@article{qu_anisotropic_2022,
	title = {Anisotropic {Fracture} of {Graphene} {Revealed} by {Surface} {Steps} on {Graphite}},
	volume = {129},
	url = {https://link.aps.org/doi/10.1103/PhysRevLett.129.026101},
	doi = {10.1103/PhysRevLett.129.026101},
	number = {2},
	urldate = {2024-11-05},
	journal = {Physical Review Letters},
	author = {Qu, Cangyu and Shi, Diwei and Chen, Li and Wu, Zhanghui and Wang, Jin and Shi, Songlin and Gao, Enlai and Xu, Zhiping and Zheng, Quanshui},
	month = jul,
	year = {2022},
	note = {Publisher: American Physical Society},
	pages = {026101},
}

@article{sutrakar_fracture_2021,
	title = {Fracture strength and fracture toughness of graphene: {MD} simulations},
	volume = {127},
	issn = {1432-0630},
	shorttitle = {Fracture strength and fracture toughness of graphene},
	url = {https://doi.org/10.1007/s00339-021-05047-x},
	doi = {10.1007/s00339-021-05047-x},
	language = {en},
	number = {12},
	urldate = {2024-11-05},
	journal = {Applied Physics A},
	author = {Sutrakar, V. K. and Javvaji, B. and Budarapu, P. R.},
	month = nov,
	year = {2021},
	pages = {949},
}

@article{hess_fracture_2016,
	title = {Fracture of perfect and defective graphene at the nanometer scale: {Is} graphene the strongest material?},
	volume = {120},
	issn = {0021-8979},
	shorttitle = {Fracture of perfect and defective graphene at the nanometer scale},
	url = {https://doi.org/10.1063/1.4962542},
	doi = {10.1063/1.4962542},
	number = {12},
	urldate = {2024-11-05},
	journal = {Journal of Applied Physics},
	author = {Hess, Peter},
	month = sep,
	year = {2016},
	pages = {124303},
}

@article{jing_effect_2012,
	title = {Effect of defects on {Young}'s modulus of graphene sheets: a molecular dynamics simulation},
	volume = {2},
	issn = {2046-2069},
	shorttitle = {Effect of defects on {Young}'s modulus of graphene sheets},
	url = {https://pubs.rsc.org/en/content/articlelanding/2012/ra/c2ra21228e},
	doi = {10.1039/C2RA21228E},
	language = {en},
	number = {24},
	urldate = {2024-11-04},
	journal = {RSC Advances},
	author = {Jing, Nuannuan and Xue, Qingzhong and Ling, Cuicui and Shan, Meixia and Zhang, Teng and Zhou, Xiaoyan and Jiao, Zhiyong},
	month = sep,
	year = {2012},
	note = {Publisher: The Royal Society of Chemistry},
	pages = {9124--9129},
}

@article{gavallas_mechanical_2023,
	title = {Mechanical properties of graphene nanoplatelets containing random structural defects},
	volume = {180},
	issn = {0167-6636},
	url = {https://www.sciencedirect.com/science/article/pii/S0167663623000571},
	doi = {10.1016/j.mechmat.2023.104611},
	urldate = {2024-10-28},
	journal = {Mechanics of Materials},
	author = {Gavallas, Panagiotis and Savvas, Dimitrios and Stefanou, George},
	month = may,
	year = {2023},
	pages = {104611},
}

@article{kumar_mechanical_2021,
	title = {Mechanical properties of graphene, defective graphene, multilayer graphene and {SiC}-graphene composites: {A} molecular dynamics study},
	volume = {620},
	issn = {0921-4526},
	shorttitle = {Mechanical properties of graphene, defective graphene, multilayer graphene and {SiC}-graphene composites},
	url = {https://www.sciencedirect.com/science/article/pii/S0921452621004245},
	doi = {10.1016/j.physb.2021.413250},
	urldate = {2025-08-13},
	journal = {Physica B: Condensed Matter},
	author = {Kumar, Yugesh and Sahoo, Sukadev and Chakraborty, Amit K.},
	month = nov,
	year = {2021},
	pages = {413250},
}

@article{torkaman-asadi_atomistic_2022,
	title = {Atomistic simulations of mechanical properties and fracture of graphene: {A} review},
	volume = {210},
	copyright = {https://www.elsevier.com/tdm/userlicense/1.0/},
	issn = {0927-0256},
	shorttitle = {Atomistic simulations of mechanical properties and fracture of graphene},
	url = {https://linkinghub.elsevier.com/retrieve/pii/S0927025622002208},
	doi = {10.1016/j.commatsci.2022.111457},
	language = {en},
	urldate = {2025-07-18},
	journal = {Computational Materials Science},
	author = {Torkaman-Asadi, M.A. and Kouchakzadeh, M.A.},
	month = jul,
	year = {2022},
	note = {Publisher: Elsevier BV},
	pages = {111457},
}

@article{akinwande_review_2017,
	title = {A review on mechanics and mechanical properties of {2D} materials—{Graphene} and beyond},
	volume = {13},
	issn = {23524316},
	url = {https://linkinghub.elsevier.com/retrieve/pii/S235243161630236X},
	doi = {10.1016/j.eml.2017.01.008},
	language = {en},
	urldate = {2025-01-24},
	journal = {Extreme Mechanics Letters},
	author = {Akinwande, Deji and Brennan, Christopher J. and Bunch, J. Scott and Egberts, Philip and Felts, Jonathan R. and Gao, Huajian and Huang, Rui and Kim, Joon-Seok and Li, Teng and Li, Yao and Liechti, Kenneth M. and Lu, Nanshu and Park, Harold S. and Reed, Evan J. and Wang, Peng and Yakobson, Boris I. and Zhang, Teng and Zhang, Yong-Wei and Zhou, Yao and Zhu, Yong},
	month = may,
	year = {2017},
	pages = {42--77},
}

@Article{LAMMPS,
  author = "A. P. Thompson and H. M. Aktulga and R. Berger and 
     D. S. Bolintineanu and W. M. Brown and P. S. Crozier and
     P. J. in 't Veld and A. Kohlmeyer and S. G. Moore and T. D. Nguyen and
     R. Shan and M. J. Stevens and J. Tranchida and C. Trott and S. J. Plimpton",
  title = "{LAMMPS} - a flexible simulation tool for
     particle-based materials modeling at the 
     atomic, meso, and continuum scales",
  journal = "Comp. Phys. Comm.",
  volume =  "271",
  pages =   "108171",
  year =    "2022",
  doi = "10.1016/j.cpc.2021.108171"
}

@article{wang_effects_2024,
	title = {The effects of grain size and fractal porosity on thermal conductivity of nano-grained graphite: {A} molecular dynamics study},
	volume = {220},
	issn = {0017-9310},
	shorttitle = {The effects of grain size and fractal porosity on thermal conductivity of nano-grained graphite},
	url = {https://www.sciencedirect.com/science/article/pii/S0017931023011754},
	doi = {10.1016/j.ijheatmasstransfer.2023.125030},
	urldate = {2024-10-28},
	journal = {International Journal of Heat and Mass Transfer},
	author = {Wang, Qian and Gui, Nan and Yang, Xingtuan and Tu, Jiyuan and Jiang, Shengyao},
	month = mar,
	year = {2024},
	pages = {125030},
}

@article{stuart_reactive_2000,
	title = {A reactive potential for hydrocarbons with intermolecular interactions},
	volume = {112},
	issn = {0021-9606},
	url = {https://doi.org/10.1063/1.481208},
	doi = {10.1063/1.481208},
	number = {14},
	urldate = {2024-11-05},
	journal = {The Journal of Chemical Physics},
	author = {Stuart, Steven J. and Tutein, Alan B. and Harrison, Judith A.},
	month = apr,
	year = {2000},
	pages = {6472--6486},
}

@article{hoover_canonical_1985,
	title = {Canonical dynamics: {Equilibrium} phase-space distributions},
	volume = {31},
	shorttitle = {Canonical dynamics},
	url = {https://link.aps.org/doi/10.1103/PhysRevA.31.1695},
	doi = {10.1103/PhysRevA.31.1695},
	number = {3},
	urldate = {2024-11-05},
	journal = {Physical Review A},
	author = {Hoover, William G.},
	month = mar,
	year = {1985},
	note = {Publisher: American Physical Society},
	pages = {1695--1697},
}

@article{ovito,
Author = {Stukowski, Alexander},
Title = {{Visualization and analysis of atomistic simulation data with OVITO-the
   Open Visualization Tool}},
Journal = {{MODELLING AND SIMULATION IN MATERIALS SCIENCE AND ENGINEERING}},
Year = {{2010}},
Volume = {{18}},
Number = {{1}},
Month = {{JAN}},
DOI = {{10.1088/0965-0393/18/1/015012}},
Article-Number = {{015012}},
ISSN = {{0965-0393}},
EISSN = {{1361-651X}},
ResearcherID-Numbers = {{Stukowski, Alexander/G-9695-2017}},
ORCID-Numbers = {{Stukowski, Alexander/0000-0001-6750-3401}},
Unique-ID = {{ISI:000272791800012}},
}

@article{jiang_mechanical_2014,
	title = {Mechanical properties of {MoS2}/graphene heterostructures},
	volume = {105},
	issn = {0003-6951, 1077-3118},
	url = {https://pubs.aip.org/apl/article/105/3/033108/931258/Mechanical-properties-of-MoS2-graphene},
	doi = {10.1063/1.4891342},
	language = {en},
	number = {3},
	urldate = {2025-02-03},
	journal = {Applied Physics Letters},
	author = {Jiang, Jin-Wu and Park, Harold S.},
	month = jul,
	year = {2014},
	pages = {033108},
}

@article{drucker_soil_1952,
	title = {Soil mechanics and plastic analysis or limit design},
	volume = {10},
	issn = {0033-569X, 1552-4485},
	url = {https://www.ams.org/qam/1952-10-02/S0033-569X-1952-48291-2/},
	doi = {10.1090/qam/48291},
	language = {en},
	number = {2},
	urldate = {2025-07-14},
	journal = {Quarterly of Applied Mathematics},
	author = {Drucker, D. C. and Prager, W.},
	month = jul,
	year = {1952},
	note = {Publisher: American Mathematical Society (AMS)},
	pages = {157--165},
}

@article{lee_measurement_2008,
	title = {Measurement of the {Elastic} {Properties} and {Intrinsic} {Strength} of {Monolayer} {Graphene}},
	volume = {321},
	issn = {0036-8075, 1095-9203},
	url = {https://www.science.org/doi/10.1126/science.1157996},
	doi = {10.1126/science.1157996},
	language = {en},
	number = {5887},
	urldate = {2025-10-09},
	journal = {Science},
	author = {Lee, Changgu and Wei, Xiaoding and Kysar, Jeffrey W. and Hone, James},
	month = jul,
	year = {2008},
	pages = {385--388},
}

@article{zhang_fracture_2015,
	title = {Fracture of graphene: a review},
	volume = {196},
	issn = {1573-2673},
	shorttitle = {Fracture of graphene},
	url = {https://doi.org/10.1007/s10704-015-0039-9},
	doi = {10.1007/s10704-015-0039-9},
	language = {en},
	number = {1},
	urldate = {2024-11-05},
	journal = {International Journal of Fracture},
	author = {Zhang, Teng and Li, Xiaoyan and Gao, Huajian},
	month = nov,
	year = {2015},
	pages = {1--31},
}

@article{sato_fracture_1987,
	title = {Fracture criteria of reactor graphite under multiaxial stesses},
	volume = {103},
	issn = {0029-5493},
	url = {https://www.sciencedirect.com/science/article/pii/0029549387903128},
	doi = {10.1016/0029-5493(87)90312-8},
	number = {3},
	urldate = {2024-10-28},
	journal = {Nuclear Engineering and Design},
	author = {Sato, S. and Awaji, H. and Kawamata, K. and Kurumada, A. and Oku, T.},
	month = sep,
	year = {1987},
	pages = {291--300},
}

@article{ely_strength_1972,
	title = {Strength of {Titania} and {Aluminum} {Silicate} {Under} {Combined} {Stresses}},
	volume = {55},
	issn = {1551-2916},
	url = {https://onlinelibrary.wiley.com/doi/abs/10.1111/j.1151-2916.1972.tb11307.x},
	doi = {10.1111/j.1151-2916.1972.tb11307.x},
	language = {en},
	number = {7},
	urldate = {2025-12-11},
	journal = {Journal of the American Ceramic Society},
	author = {Ely, Richard E.},
	year = {1972},
	note = {\_eprint: https://ceramics.onlinelibrary.wiley.com/doi/pdf/10.1111/j.1151-2916.1972.tb11307.x},
	pages = {347--350},
}

@article{broutman_effects_1970,
	title = {Effects of {Combined} {Stresses} on {Fracture} of {Alumina} and {Graphite}},
	volume = {53},
	issn = {1551-2916},
	url = {https://onlinelibrary.wiley.com/doi/abs/10.1111/j.1151-2916.1970.tb12034.x},
	doi = {10.1111/j.1151-2916.1970.tb12034.x},
	language = {en},
	number = {12},
	urldate = {2025-12-11},
	journal = {Journal of the American Ceramic Society},
	author = {Broutman, L. J. and Krishnakumar, S. M. and Mallick, P. K.},
	year = {1970},
	note = {\_eprint: https://ceramics.onlinelibrary.wiley.com/doi/pdf/10.1111/j.1151-2916.1970.tb12034.x},
	pages = {649--654},
}

@book{bishop_pattern_2006,
	address = {New York},
	series = {Information science and statistics},
	title = {Pattern recognition and machine learning},
	isbn = {978-0-387-31073-2},
	language = {en},
	publisher = {Springer},
	author = {Bishop, Christopher M.},
	year = {2006},
}

@book{burnham2002,
  title={Model Selection and Multimodel Inference: A Practical Information-Theoretic Approach},
  author={Burnham, Kenneth P. and Anderson, David R.},
  year={2002},
  publisher={Springer}
}

@book{jolliffe_principal_2002,
	address = {New York},
	edition = {2nd ed},
	series = {Springer series in statistics},
	title = {Principal component analysis},
	isbn = {978-0-387-95442-4},
	language = {eng},
	publisher = {Springer},
	author = {Jolliffe, Ian T.},
	year = {2002},
}
\end{document}